\documentclass[journal,10pt]{IEEEtran}
\usepackage{amsfonts}
\usepackage{amssymb}
\usepackage{eurosym}
\usepackage{cite}
\usepackage{graphicx}
\usepackage{epstopdf}
\usepackage{amsmath}
\usepackage{tikz,lipsum}
\usepackage{caption}
\usepackage{float}
\usepackage[T1]{fontenc}
\usepackage{amsthm}
\usepackage{mathrsfs}
\usepackage{slashbox}
\usepackage{mathrsfs}
\usepackage{color}
\usepackage{geometry}

\IEEEoverridecommandlockouts
\newcounter{mytempeqncnt}
\newtheorem{remark}{Remark}

\newtheorem{lemma}{Lemma}
\newtheorem{proposition}{Proposition}
\newtheorem{corollary}{Corollary}

\begin{document}

\title{Analysis of Asymmetric Dual-Hop Energy Harvesting-Based Wireless
Communication Systems in Mixed Fading Environments}
\author{Elmehdi~Illi,~\IEEEmembership{Member, IEEE,} Faissal~El~Bouanani,~%
\IEEEmembership{Senior Member, IEEE,} Paschalis C. Sofotasios,
\IEEEmembership{Senior Member,
IEEE,} Sami Muhaidat, \IEEEmembership{Senior
Member, IEEE,} Daniel Benevides da Costa,
\IEEEmembership{Senior Member,
IEEE,} Fouad~Ayoub,~\IEEEmembership{Member,~IEEE,}~and~Ala Al-Fuqaha,~%
\IEEEmembership{Senior Member,~IEEE}
\IEEEcompsocitemizethanks{\IEEEcompsocthanksitem E. Illi and F. El Bouanani are with ENSIAS College of Engineering, Mohammed V University, Rabat, Morocco (e-mails: \{elmehdi.illi, f.elbouanani\}@um5s.net.ma).\protect \and
\IEEEcompsocthanksitem P. C. Sofotasios and S. Muhaidat are with the Center for Cyber-Physical Systems, Department of Electrical and Computer Engineering, Khalifa University, PO Box 127788, Abu Dhabi, UAE (e-mails: \{p.sofotasios, muhaidat\}@ieee.org).\protect \and
\IEEEcompsocthanksitem D. B. da Costa is with the Department of Computer Engineering, Federal University of Cear\'a (UFC), Sobral-CE, Brazil (e-mail: danielbcosta@ieee.org).\protect \and
\IEEEcompsocthanksitem F. Ayoub is with CRMEF, Kenitra, Morocco (e-mail: ayoub@crmefk.ma).\protect \and
\IEEEcompsocthanksitem A. Al-Fuqaha is with the Department of Computer Science, Western Michigan University, Kalamazoo, MI 49008, USA
(e-mail: ala.al-fuqaha@wmich.edu).
}}
\maketitle

\begin{abstract}
In this paper, the performance of a dual-hop energy harvesting-based
fixed-gain amplify-and-forward (AF) relaying communication system, subject
to fading impairments, is investigated. We consider a source node ($S$)
communicating with a destination node ($D$) through a fixed distant relay ($R
$), which harvests energy from its received signals and uses it to amplify
and forward the received signals to $D$. Power-splitting (PS) and
time-switching (TS) schemes are considered in the analysis for energy
harvesting. The $S$-$R$ and $R$-$D$ hops are modeled by the Nakagami-$m$ and
$\alpha $-$\mu $ fading models, respectively. Closed-form expressions for
the statistical properties of the end-to-end signal-to-noise ratio (SNR) are
derived, based on which novel closed-form expressions for the average symbol
error rate (ASER) as well as average channel capacity (ACC) considering four
adaptive transmission policies are derived. The derived expressions are
validated through Monte-Carlo simulations.
\end{abstract}



\IEEEpeerreviewmaketitle

\section{Introduction}

The ever-increasing demand for broadband communication systems and the
growing number of connected devices are the driving forces for the evolution
of wireless technologies in the past decades \cite{5G}. With the emergence
of new paradigms such as the Internet of Things (IoT) as well as
machine-to-machine (M2M) communication, the next generation of wireless
networks are envisioned to catalyze the deployment of new technologies and
services, such as remote healthcare, surveillance, and transportation \cite%
{m2m}.

In addition to the need for higher data rates and expansion of network
coverage, the maximization of the energy efficiency is among the most
critical challenges for the fifth generation (5G) of wireless networks and
beyond \cite{zhang2017}, where the power consumption and battery lifetime of
wireless nodes are of paramount importance \cite{5G}, \cite{seddik}.
Although energy resources (e.g., battery) are limited, the connected devices
in 5G\ systems are envisaged to operate in multiple spectrum bands, as well
as providing a real-time processing \cite{hossain}. Additionally, M2M
devices and wireless sensors are typically deployed in difficult-to-reach
areas, e.g., structural health monitoring and mine tunnels \cite{hossain},
\cite{pan1}, making the battery recharging or replacement impractical in
most cases.

Recently, energy harvesting (EH) has emerged as an attractive solution that
is envisioned to provide a greener and safer energy supply to
self-sustainable wireless commutations \cite{sudevalayam}. Radio-frequency
energy harvesting (RF-EH) was proposed recently as a promising solution to
provide perpetual energy replenishment for wireless networks. RF-EH is
realized by allowing wireless devices, equipped with dedicated EH circuits,
to harvest energy from either ambient RF signals or dedicated RF sources
\cite{kamaliwsn}, \cite{danieleh}. In M2M communications\ and wireless
sensor networks (WSN), terminal nodes harvest energy from either access
points or a dedicated power source/base stations \cite{5G}. RF-EH can be
categorized into two main strategies, namely, wireless-powered
communications (WPC) \cite{samiref} and simultaneous wireless information
and power transfer (SWIPT), which has been shown to provide noticeable gains
in terms of power and spectral efficiencies by enabling simultaneous
information processing and wireless power transfer \cite{pan1}. In practical
scenarios, it is difficult to perform information decoding and energy
harvesting. Accordingly, two practical system designs, namely time-switching
(TS), and power-splitting (PS), were proposed \cite{pan1}. In the former,
the receiver switches over time between information decoding and EH, while
in the latter, the received power is split into two streams, one for EH and
the other for information processing \cite{miso}.

On the other hand, RF communications are often impaired due to multipath
fading and shadowing random phenomena, caused by the presence of reflectors,
scatterers, and obstacles \cite{alouinisimon}. Within this context, several
existing distributions were introduced, supported by field test
measurements, to accurately describe the statistics of the signal variations
due to fading, namely Rayleigh for a non-line of sight (NLOS)-based link,
Rician for a LOS-based communication, as well as Nakagami-$m$ and Weibull
distributions for urban outdoor environments. In \cite{yacoub}, $\alpha $-$%
\mu $ model was proposed as an alternative solution for modeling the fading
amplitude in a mobile radio channel, since it includes a vast majority of
the well-known fading distributions, namely Rayleigh, Weibull, and Nakagami-$%
m,$ as special cases.

Multihop relaying, where information is communicated between two terminals
(nodes) over multiple hops, has been extensively investigated in the
literature. This multihop approach realizes several key advantages as
compared to single-hop scenario, e.g., lower power consumption and better
throughput. A variant of multihop relaying, known as cooperative diversity,
or cooperative communications, has emerged recently as a promising approach
to increase spectral and power efficiencies, network coverage, and reduce
outage probability, mostly used in infrastructure-less based networks. In
cooperative communications, the relays process signals overheard from the
source terminal and re-transmit them toward a destination. Two common
relaying techniques are decode-and-forward (DF) and amplify-and-forward (AF)
\cite{relay}. In DF relaying, the relay terminal decodes a received signal
and then re-encodes it (possibly using a different codebook) for
transmission to a destination. With AF relaying, the relay terminal
re-transmits a scaled version of the received signal without any attempt to
decode it \cite{relay}. Also, in scenarios, the cooperating nodes are
typically located at different locations, and, therefore, asymmetric
channels are often experienced \cite{paschalis}. Therefore, several research
studies investigated the performance analysis of dual-hop transmission
systems under asymmetric fading environments, for AF and DF relaying.
Specifically, in \cite{kapucu}, the authors investigated the performance of
DF relaying systems in mixed fading environments, which were modeled by
Rayleigh/Generalized-Gamma fading channels. The authors in \cite{surawera}
examined the performance of a dual-hop system in mixed fading environments
subject to Rayleigh/Rician scenarios. In \cite{arqrelay}, the analysis was
carried out assuming a dual-hop DF relaying protocol with HARQ
retransmission scheme in Rayleigh/Rician environments, while the work in
\cite{paschalis} dealt with the analysis of a full-duplex DF relaying system
over generalized fading channels. Finally, the work in \cite{jaya} focused
on the analysis of a multiple-input multiple-output (MIMO) relay network
with dual-hop AF relaying, over asymmetric fading channels.

Within the context of EH, the performance of multi-hop EH-based wireless
communication systems has been widely investigated in the literature. In
\cite{ref1}, the authors dealt with the bit error rate performance of a
dual-hop EH-based system, by considering the AF and DF protocols\ and taking
into account TS and PS, over Nakagami-$m$ fading channels. In \cite{ref2},
the performance of a dual-hop full-duplex EH-based system is analyzed, by
considering both AF\ and DF\ protocols, while \cite{ref3} dealt with the
analysis of a multi-hop cognitive-radio EH-based network. The works in \cite%
{danieleh} and \cite{ref5} analyzed the performance of a dual-hop
multi-relay communication scheme and a MIMO dual-hop EH-based communication
system, respectively.

Although the previous works added new insights, they have mainly focused on:
outage probability, error rate, and capacity analysis. Furthermore, sporadic
results have been reported on the performance of dual-hop EH-based
communication systems subject to asymmetric fading conditions. To the best
of the authors' knowledge, the\ performance analysis of asymmetric dual-hop
EH-based communication systems, including the average symbol error rate
(ASER)\ and the average channel capacity (ACC) over two distinct adaptive
transmission policies, has not been addressed yet in the open literature.
Owing to this fact, this work aims to fill this gap by investigating the
ASER and capacity performance of EH-based dual-hop fixed-gain AF relaying
systems over four different adaptive transmit policies, namely, optimal
power adaptation (ORA), optimal power and rate adaptation (OPRA), channel
inversion with fixed rate (CIFR), and truncated channel inversion with fixed
rate (TCIFR). In particular, asymmetric fading conditions are assumed over
the two hops, with Nakagami-$m$ and $\alpha $-$\mu $ fading conditions being
taken into account, for the first and second hops, respectively. Both TS and
PS are considered in the analysis. Overall, the main contributions of this
paper can be summarized as follows:

\begin{itemize}
\item Novel generalized closed-form expressions for the cumulative
distribution function (CDF) and probability density function (PDF) of the
end-to-end SNR of the considered system are derived, for both TS\ and PS
schemes.

\item Based on the statistical properties, novel closed-form expressions for
the ASER\ for various modulation schemes and the ACC over four adaptive
policies are derived.
\end{itemize}

The remainder of this paper is organized as follows. Section II describes
the considered system and channel models, while Section III is dedicated to
the statistical properties of the end-to-end SNR, namely the CDF and PDF.
Section IV focuses on the performance analysis of the considered system over
various adaptive policies. Numerical results and discussions are presented
in Section V, while conclusions are drawn in Section VI.

\section{Channel and System Models}

\begin{figure}[tbp]
\begin{center}
\includegraphics[scale=.4]{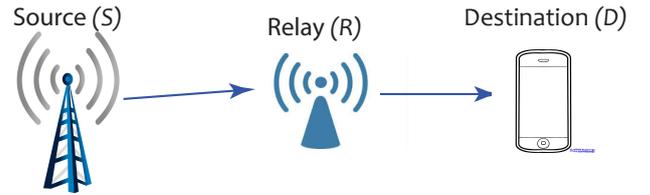}\vspace{-7cm}
\end{center}
\caption{System model}
\label{sysmodd}
\end{figure}
We consider a dual-hop EH-based wireless communication system with a
fixed-gain AF relay, operating with both TS and PS\ schemes, as shown in
Fig. \ref{sysmodd} \footnote{%
Nakagami-$m$ fading model has been widely advocated in literature for
providing a good agreement with urban land-mobile communication scenarios
\cite{alouinisimon}, \cite{rubio}. On the other hand, $\alpha $-$\mu $ model
can generalize fading scenarios for the same abovementioned environment, as
well as for modeling ionospheric scintillation as shown in \cite{moraes1}
and \cite{moraes2}.}. A source terminal $S$ communicates with the
destination $D$ through a fixed relay node $R$ over two time slots. For\ the
TS$\ $scheme, $S$ transmits the information signal to $R$\ during the first
part of the first time slot, and then transmits energy signal during the
remaining part of it, while for PS, the power is divided at the relay into
two portions, one for information decoding and the other for EH. The relay $R
$ uses the energy harvested during Phase-1 to amplify the received signal
and forward it to $D$. {It is assumed that no direct link exists between $S$
and $D$. We further assume asymmetric fading condition over the $S$-$R$ and $%
R$-$D$ hops. In particular, the $S$-$R$ link is subject to Nakagami-$m$
fading channel, while the $R$-$D$ channel undergoes $\alpha $-$\mu $ fading.
}

\subsection{Phase I: $S$-$R$ hop}

\subsubsection{\protect\bigskip Information Signal Transmission}

\begin{itemize}
\item \textbf{TS protocol}: The source node sends during the first $%
(1-\varepsilon )T_{0}$ seconds the information signal $x_{s}$ to $R$, where $%
\varepsilon $ stands for the time portion dedicated to the EH, and $T_{0}$
is the time slot dedicated to the $S$-$R$ communication. The received
instantaneous SNR at the relay can be expressed as \cite{alouinisimon}
\begin{equation}
\gamma _{1}^{(TS)}=\frac{P_{S}}{d_{1}^{\delta }N_{1}}\left\vert
h_{1}\right\vert ^{2},  \label{gam1}
\end{equation}%
in which $P_{S}$ denotes the transmit power, $d_{R}$ is the $S$-$R$
distance, $\delta $ is the path-loss exponent, $h_{R}$ denotes the $S$-$R$
channel gain with average fading power $\Omega _{1}=\mathbb{E}_{\left\vert
h_{1}\right\vert }\left[ \left\vert h_{1}\right\vert ^{2}\right] $, $\mathbb{%
E}_{X}\mathbb{[}.\mathbb{]}$ denotes the expected value with respect to the
random variable $X,$ and $N_{1}$ denotes additive white Gaussian noise
(AWGN) power.

\item \textbf{PS protocol: }$S$ sends during the first time slot $T_{0}$ the
information signal $x$ with power $P_{S}.$ The relay splits the power of the
received signal into two parts: A portion $1-\varrho $ of the received power
at the relay is dedicated to information decoding, where $0<\varrho <1$ is
the power portion dedicated to EH. The received instantaneous SNR at the
relay assuming PS\ scheme can be expressed as
\begin{equation}
\gamma _{1}^{(PS)}=\frac{\left( 1-\varrho \right) P_{S}}{d_{1}^{\delta }N_{1}%
}\left\vert h_{1}\right\vert ^{2}.
\end{equation}

Consequently, the SNR\ $\gamma _{1}$ can be generalized for both schemes as
follows%
\begin{equation}
\gamma _{1}=\frac{\left( 1-\varsigma \right) P_{S}}{d_{1}^{\delta }N_{1}}%
\left\vert h_{1}\right\vert ^{2}.  \label{snr1ps}
\end{equation}%
with an average value
\begin{equation}
\overline{\gamma }_{1}=\frac{\left( 1-\varsigma \right) P_{S}}{d_{1}^{\delta
}N_{1}}\Omega _{1},
\end{equation}%
and $\varsigma $ equals either $0$ for TS, and $\varrho $ for PS. In this
work, Nakagami-$m$ fading model is considered for modelling the channel gain
distribution of the $S$-$R$ hop. As a consequence, the received SNR$\ $at
the relay is Gamma-distributed with PDF/CDF\ expressions given as \cite%
{alouinisimon}%
\begin{eqnarray}
f_{\gamma _{1}}\left( z\right)  &=&\left( \frac{m_{1}}{\overline{\gamma }_{1}%
}\right) ^{m}\frac{z^{m_{1}-1}}{\Gamma \left( m_{1}\right) }\exp \left( -%
\frac{m_{1}}{\overline{\gamma }_{1}}z\right) ,  \label{pdfsx} \\
F_{\gamma _{1}}\left( z\right)  &=&\frac{\gamma _{\text{inc}}\left( m_{1},%
\frac{m_{1}}{\overline{\gamma }_{1}}z\right) }{\Gamma \left( m_{1}\right) },
\label{cdfsx}
\end{eqnarray}%
where $m_{1}$ stands for the Nakagami-$m$ fading parameter, $\Gamma \left(
.\right) $ and $\gamma _{\text{inc}}\left( .,.\right) $ denote the Gamma and
the lower-incomplete Gamma functions, respectively \cite[Eqs. (8.310.1),
(8.350.1)]{integrals}.
\end{itemize}

\subsubsection{Energy Harvesting}

For the TS scheme,\ the information is transmitted from $S$ to $D$ during a
total time of $T_{0}$. Under this scheme, $\varepsilon $ represents the
portion of $T_{0}$ in which the relay harvests energy from the source
signal, where $0<\varepsilon <1$. Subsequently, data transmission is carried
out in the remaining block transmission time. However, for the PS\ scheme, a
portion $\varrho $ of the received power at the receiver is harvested by the
relay. As a result, the collected energy by the relay for both schemes can
be written as

\begin{equation}
E_{R}=\frac{\theta \kappa T_{0}P_{S}\left\vert h_{1}\right\vert ^{2}}{%
d_{1}^{\delta }},  \label{nrj}
\end{equation}%
where $\kappa $ denotes $\varepsilon $ for TS and $\varrho $ for PS, and $%
\theta $ stands for the conversion efficiency of the relay's energy
harvester.

\bigskip The relay $R$ uses the harvested energy to amplify the received
signal and to transmit it to $D$ during the time slot $T_{1}.$ Henceforth, a
finite battery storage model is assumed, where the harvested energy at the
relay, $E_{R},$ can either be less than the relay battery capacity, or
exceeds it. In the former case, the relay node communicates with $D$ with a
power $P_{E}=\frac{E_{R}}{T_{1}}$, while in the latter one, the $R$-$D$
communication is ensured with the whole available power at the battery;
i.e., $P_{B}=\frac{B_{R}}{T_{1}},$ with $B_{R}$ being the battery capacity
in mAh$\times $V. Thus, the relay transmit power $P_{R}$ can be expressed as%
\begin{equation}
P_{R}=\left\{
\begin{array}{c}
P_{E};\text{ if }E_{R}<B_{R} \\
P_{B};\text{ if }E_{R}\geq B_{R}%
\end{array}%
\right. .  \label{pwr}
\end{equation}

It can be easily shown that the PDF and CDF of $P_{E}$ are given by
\begin{eqnarray}
f_{P_{E}}\left( y\right) &=&\Psi ^{m_{1}}\frac{y^{m_{1}-1}}{\Gamma \left(
m_{1}\right) }\exp \left( -\Psi y\right) ,  \label{pdfpwr} \\
F_{P_{E}}\left( y\right) &=&\frac{\gamma _{\text{inc}}\left( m_{1},\Psi
y\right) }{\Gamma \left( m_{1}\right) },  \label{cdfpwr}
\end{eqnarray}%
where
\begin{equation}
\Psi =\frac{m_{1}T_{1}d_{1}^{\delta }}{\theta \kappa T_{0}P_{s}\Omega _{1}}.
\label{psii}
\end{equation}

\subsection{Phase II: $R$-$D$ hop}

\subsubsection{$E_{R}<B_{R}$}

During the second time slot $T_{1}$, the relay amplifies the signal received
from $S$ by a fixed gain, and forwards it to $D$ using the harvested power $%
P_{E}.$

The received SNR at $D$ over the $R$-$D$ link is given by
\begin{eqnarray}
\gamma _{2}^{(1)} &=&\frac{P_{E}\left\vert h_{2}\right\vert ^{2}}{%
d_{2}^{\delta }N_{2}}  \notag \\
&=&P_{E}\Upsilon _{2}.  \label{gam2}
\end{eqnarray}

Here, the $R$-$D$ fading coefficient is assumed to follow an $\alpha $-$\mu $
distribution. Consequently, the SNR $\Upsilon _{2}$ is also $\alpha $-$\mu $
distributed, whose PDF is given by \cite{yacoub}
\begin{equation}
f_{\Upsilon _{2}}(z)=\frac{\alpha _{2}\mu _{2}^{\mu _{2}}}{2\Gamma (\mu _{2})%
\overline{\Upsilon }_{2}}\left( \frac{z}{\overline{\Upsilon }_{2}}\right) ^{%
\frac{\alpha _{2}\mu _{2}}{2}-1}e^{-\mu _{2}\left( \frac{z}{\overline{%
\Upsilon }_{2}}\right) ^{\frac{\alpha _{2}}{2}}},  \label{gggt}
\end{equation}%
with average value $\overline{\Upsilon }_{2}=\frac{\Omega _{2}}{%
d_{2}^{\delta }N_{2}},$ where $\alpha _{2}$ and $\mu _{2}$ denote the two
physical fading parameters that reflect the nonlinearity and clustering,
respectively, and $\overline{\gamma }_{2}^{(1)}$ is the average $R$-$D$ SNR\
of this first case, given as
\begin{equation}
\overline{\gamma }_{2}^{(1)}=\mathbb{E}_{P_{E}}\left[ P_{E}\right] \overline{%
\Upsilon }_{2},
\end{equation}%
where $\mathbb{E}_{P_{E}}\left[ P_{E}\right] =\frac{\theta \kappa
T_{0}P_{s}\Omega _{1}}{T_{1}d_{1}^{\delta }}.$

\subsubsection{$E_{R}\geq B_{R}$}

In this scenario, the relay $R$ forwards the information signal to $D,$
after amplifying it with a fixed gain with power $P_{B}=\frac{B_{R}}{T_{1}}.$
Consequently, the received SNR at $D$ is expressed as%
\begin{equation}
\gamma _{2}^{(2)}=P_{B}\Upsilon _{2},  \label{snr22}
\end{equation}%
with an average value%
\begin{equation*}
\overline{\gamma }_{2}^{(2)}=P_{B}\overline{\Upsilon }_{2}.
\end{equation*}

\section{Statistical properties}

In this section, the CDF expression in closed-form for the end-to-end SNR is
derived.

Since the relay node performs fixed-gain AF relaying, the total received
SNR\ at $D$ is expressed as \cite{relay}
\begin{equation}
\gamma _{eq}=\frac{\gamma _{1}\gamma _{2}}{\gamma _{2}+C},  \label{mn}
\end{equation}%
with $C$ being a fixed-gain relaying constant.

\subsection{Cumulative Distribution Function}

\begin{proposition}
\bigskip The CDF\ expression of the end-to-end SNR\ for the considered
system is expressed as%
\begin{equation}
F_{\gamma _{eq}}(z)=F_{\gamma _{eq}^{(1)}}(z)F_{P_{E}}\left( P_{B}\right)
+F_{\gamma _{eq}^{(2)}}(z)F_{P_{E}}^{c}\left( P_{B}\right) ,
\label{cdffinal}
\end{equation}%
with
\begin{equation}
\gamma _{eq}^{(i)}=\frac{\gamma _{1}\gamma _{2}^{(i)}}{\gamma _{2}^{(i)}+C},
\label{snreqi}
\end{equation}%
$F_{X}^{c}\left( .\right) $ is the complementary CDF\ of $X$, and%
\begin{align}
F_{\gamma _{eq}^{(1)}}(z)& =1-\frac{\alpha _{2}}{2\Gamma (m_{1})\Gamma (\mu
_{2})}e^{-\frac{m_{1}}{\overline{\gamma }_{1}}z}\sum_{n=0}^{m_{1}-1}%
\sum_{p=0}^{n}\left(
\begin{array}{c}
n \\
p%
\end{array}%
\right)  \notag \\
& \times \frac{\left( \frac{m_{1}}{\overline{\gamma }_{1}}z\right) ^{n-p}}{n!%
}H_{0,3}^{3,0}\left( \mu _{2}\left( \frac{m_{1}^{2}Cz}{\overline{\gamma }_{1}%
\overline{\gamma }_{2}^{(1)}}\right) ^{\frac{\alpha _{2}}{2}}\left\vert
\begin{array}{c}
-;- \\
\Delta _{1};-%
\end{array}%
\right. \right) ,  \label{cdffinal1}
\end{align}%
and%
\begin{align}
F_{\gamma _{eq}^{(2)}}\left( z\right) & =1-\frac{\alpha _{2}}{2\Gamma (\mu
_{2})}e^{-\frac{m_{1}}{\overline{\gamma }_{1}}z}\sum_{n=0}^{m_{1}-1}%
\sum_{p=0}^{n}\left(
\begin{array}{c}
n \\
p%
\end{array}%
\right) \frac{\left( \frac{m_{1}}{\overline{\gamma }_{1}}z\right) ^{n-p}}{n!}
\notag \\
& \times H_{0,2}^{2,0}\left( \mu _{2}\left( \frac{m_{1}Cz}{\overline{\gamma }%
_{1}\overline{\gamma }_{2}^{(2)}}\right) ^{\frac{\alpha _{2}}{2}}\left\vert
\begin{array}{c}
-;- \\
\Delta _{2};-%
\end{array}%
\right. \right) ,  \label{cdffinal2}
\end{align}%
where $H_{p,q}^{m,n}\left( z\left\vert
\begin{array}{c}
(a_{i},A_{i})_{i=1:p} \\
(b_{k},B_{k})_{k=1:q}%
\end{array}%
\right. \right) $, $p\geq n,$ and $q\geq m,$ denotes the Fox's $H$-Function
\cite{mathai}, $\Delta _{1}=(\mu _{2},1),\left( p,\frac{\alpha _{2}}{2}%
\right) ,\left( m_{1}+\mu _{2}\left( 1-\frac{\alpha _{2}}{2}\right) ,\frac{%
\alpha _{2}}{2}\right) ,$ and $\Delta _{2}=(\mu _{2},1),\left( p,\frac{%
\alpha _{2}}{2}\right) .$%
\begin{IEEEproof}
The proof is provided in Appendix A.
\end{IEEEproof}
\end{proposition}

\subsection{Probability Density Function}

\begin{corollary}
The PDF\ of the end-to-end SNR of the considered system can be retrieved
readily by differentiating (\ref{cdffinal}) yielding%
\begin{align}
f_{\gamma _{eq}}(z)& =\frac{\alpha _{2}e^{-\frac{m_{1}}{\overline{\gamma }%
_{1}}z}}{2\Gamma (\mu _{2})}\sum_{n=0}^{m_{1}-1}\sum_{p=0}^{n}\left(
\begin{array}{c}
n \\
p%
\end{array}%
\right) \frac{\left( \frac{m_{1}}{\overline{\gamma }_{1}}z\right) ^{n-p}}{n!}
\notag \\
& \times \left( \frac{F_{P_{E}}\left( P_{B}\right) }{\Gamma (m_{1})}%
U_{1}+F_{P_{E}}^{c}\left( P_{B}\right) U_{2}\right) ,  \label{pdffinal}
\end{align}%
with
\begin{equation}
U_{i}=\left[
\begin{array}{l}
\left( \frac{m_{1}}{\overline{\gamma }_{1}}-\frac{n-p}{z}\right)
H_{0,3-i+1}^{3-i+1,0}\left( \varkappa z^{\frac{\alpha _{2}}{2}}\left\vert
\begin{array}{c}
-;- \\
\Delta _{i};-%
\end{array}%
\right. \right) \\
-\frac{\alpha _{2}}{2z}H_{1,4-i+1}^{3-i+1,1}\left( \varkappa z^{\frac{\alpha
_{2}}{2}}\left\vert
\begin{array}{c}
(0,1);- \\
\Delta _{i};(1,1)%
\end{array}%
\right. \right)%
\end{array}%
\right] ,i=1,2.
\end{equation}%
with $\varkappa =\mu _{2}\left( \frac{m_{1}^{2-i+1}Cz}{\overline{\gamma }%
_{2}^{(i)}\overline{\gamma }_{1}}\right) ^{\frac{\alpha _{2}}{2}}.$%
\begin{IEEEproof}
The proof is provided in Appendix B.
\end{IEEEproof}
\end{corollary}

\section{Performance Evaluation}

In this section, closed-form expressions for the ASER for different
modulation techniques, and the ACC of four adaptive transmission policies,
namely ORA, OPRA, CIFR, and TCIFR, are derived.

\begin{lemma}
\label{laplace2} Let $x\geq 0$ be a positive real number. It follows that
\begin{align}
\Xi (x)& =\int_{x}^{\infty }z^{\omega -1}e^{-\lambda z}H_{p,q}^{m,n}\left(
az^{v}\left\vert
\begin{array}{c}
(a_{i},A_{i})_{j=1:p} \\
(b_{k},B_{k})_{k=1:q}%
\end{array}%
\right. \right) dz  \notag \\
& =\lambda ^{-\omega }M_{p+1,q}^{m,n+1}\left( a\lambda ^{-v}\left\vert
\begin{array}{c}
w,(a_{i},A_{i},0)_{i=2:p+1} \\
(b_{k},B_{k},0)_{k=1:q}%
\end{array}%
\right. \right) ,  \label{mypsi}
\end{align}%
where $w=(1-\omega ,v;\lambda x),$ $M_{p,q}^{m,n}\left( z\left\vert
\begin{array}{c}
\left( a_{i},A_{i},\alpha _{i}\right) _{i=1:p} \\
\left( b_{k},B_{k},\beta _{k}\right) _{k=1:q}%
\end{array}%
\right. \right) $ denotes the generalized incomplete-upper Fox's $H$%
-Function \cite{incomplete}.
\begin{IEEEproof}
The proof is provided in Appendix C.
\end{IEEEproof}
\end{lemma}

It is noteworthy that the Fox's $H$-Function and the generalized
incomplete-upper Fox's $H$-Function can be implemented efficiently in most
popular computer software, such as Matlab or Mathematica \cite{incomplete}.

\subsection{\protect\bigskip Average Symbol Error Rate}

The ASER\ is a common performance metric for evaluating the communication
reliability over fading channels. For a communication system subject to
random fading, it is defined as the statistical average value of the
instantaneous symbol error rate. For a variety of signaling/modulation
techniques, the ASER is defined as%
\begin{eqnarray}
\overline{P}_{se} &=&\mathbb{E}_{\gamma }[P_{se}(\gamma )]  \notag \\
&=&\int_{0}^{\infty }P_{se}(\gamma )f_{\gamma }(\gamma )d\gamma .
\label{aser1}
\end{eqnarray}

For various modulation types, the ASER\ is given by
\begin{equation}
\overline{P}_{se}=\rho \int_{0}^{\infty }\mathrm{erfc}(\sqrt{\gamma \theta }%
)f_{\gamma }(\gamma )d\gamma ,  \label{aser2}
\end{equation}%
where $\rho $ and $\theta $ are two modulation-dependant parameters \cite%
{illi}, and $\mathrm{erfc}\left( .\right) $ stands for the complementary
Gaussian error function \cite[Eqs. (8.250.1, 8.250.4)]{integrals}.

\begin{proposition}
The ASER\ of the considered communication system for a variety of modulation
schemes is expressed as%
\begin{equation}
\overline{P}_{se}=\rho \left[
\begin{array}{l}
1-\frac{\alpha _{2}\sqrt{\frac{\tau }{\sigma \pi }}}{2\Gamma (m_{1})\Gamma
(\mu _{2})}\sum\limits_{n=0}^{m_{1}-1}\sum\limits_{p=0}^{n}\left(
\begin{array}{c}
n \\
p%
\end{array}%
\right) \frac{\left( \frac{m_{1}}{\overline{\gamma }_{1}\sigma }\right)
^{n-p}}{n!} \\
\times \left[ F_{P_{E}}\left( P_{B}\right) V_{1}+F_{P_{E}}^{c}\left(
P_{B}\right) \Gamma (m_{1})V_{2}\right]%
\end{array}%
\right] ,  \label{aserfinal}
\end{equation}%
where%
\begin{equation}
V_{i}=H_{1,3-i+1}^{3-i+1,1}\left( \varkappa \left\vert
\begin{array}{c}
\left( p-n+\frac{1}{2},\frac{\alpha _{2}}{2}\right) ;- \\
\Delta _{i};-%
\end{array}%
\right. \right) ,i=1,2,
\end{equation}%
\begin{equation}
\sigma =\frac{m_{1}}{\overline{\gamma }_{1}}+\tau .
\end{equation}
\end{proposition}

\begin{IEEEproof}
The ASER\ expression in (\ref{aser2}) can be expressed using the CDF\ as%
\begin{equation}
\overline{P}_{se}=\rho \int_{0}^{\infty }z^{\frac{-1}{2}}e^{-\tau
z}F_{\gamma _{eq}}(\gamma )d\gamma .  \label{aser3}
\end{equation}

By involving the CDF\ in (\ref{cdffinal}) into (\ref{aser3}), and
making use of the identity \cite[Eq. (2.19)]{mathai} alongside with some
algebraic manipulations, (\ref{aserfinal}) is attained.
\end{IEEEproof}

\subsection{Optimal rate adaptation policy}

The bandwidth-normalized ACC under constant transmission power, namely ORA
policy, is often known as Shannon capacity. By definition, it is expressed
as \cite{alouinisimon}%
\begin{equation}
\overline{C}_{ORA}=\int_{0}^{\infty }\log _{2}\left( 1+z\right) f_{\gamma
_{eq}}(z)dz,
\end{equation}

Alternatively, it can be expressed in terms of the complementary CDF\ of the
SNR as
\begin{equation}
\overline{C}_{ORA}=\frac{1}{\ln (2)}\int_{0}^{\infty }\frac{F_{\gamma
_{eq}}^{c}(z)}{1+z}dz.  \label{ora1}
\end{equation}

\begin{proposition}
\begin{figure*}[t]
{\normalsize 
\setcounter{mytempeqncnt}{\value{equation}}
\setcounter{equation}{31} }
\par
\begin{equation}
\overline{C}_{ORA}=\frac{\alpha _{2}\overline{\gamma }_{1}}{2\ln (2)\Gamma
(\mu _{2})m_{1}}\sum_{n=0}^{m_{1}-1}\sum_{p=0}^{n}\left(
\begin{array}{c}
n \\
p%
\end{array}%
\right) \frac{1}{n!}\left[
\begin{array}{c}
\frac{F_{P_{E}}\left( P_{B}\right) }{\Gamma (m_{1})}%
H_{1,0:1,1:0,3}^{0,1:1,1:3,0}\left( \frac{\overline{\gamma }_{1}}{m_{1}};\mu
_{2}\left( \frac{m_{1}C}{\overline{\gamma }_{2}^{(1)}}\right) ^{\frac{\alpha
_{2}}{2}}\left\vert \Lambda _{1}\right. \right) \\
+F_{P_{E}}^{c}\left( P_{B}\right) H_{1,0:1,1:0,2}^{0,1:1,1:2,0}\left( \frac{%
\overline{\gamma }_{1}}{m_{1}};\mu _{2}\left( \frac{C}{\overline{\gamma }%
_{2}^{(2)}}\right) ^{\frac{\alpha _{2}}{2}}\left\vert \Lambda _{2}\right.
\right)%
\end{array}%
\right] .  \label{orafinal}
\end{equation}%
\par
{\normalsize 
\hrulefill 
\vspace*{4pt} }
\end{figure*}
The ACC\ under ORA\ policy of the considered AF\ dual-hop system is
expressed in (\ref{orafinal}) at the top of the next page where $\Lambda
_{k}=\left(
\begin{array}{c}
(-n+p,1,\frac{\alpha _{2}}{2});-:(0,1);-:-;- \\
-;-:(0,1);-:\Delta _{k};-%
\end{array}%
\right) _{k=1,2},$ $%
H_{p_{1},q_{1};p_{2},q_{2};p_{3},q_{3}}^{0,n_{1};m_{2},n_{2};m_{3},n_{3}}%
\left( x,y\left\vert \Xi \right. \right) ,$ with $\Xi =%
\begin{array}{c}
(a_{i},\alpha
_{i};A_{i})_{i=1:p_{1}};(c_{i},C_{i})_{i=1:p_{2}};(e_{i},E_{i})_{i=1:p_{3}}
\\
(b_{k},\beta
_{k};B_{k})_{k=1:q_{1}};(d_{k},D_{k})_{k=1:q_{2}};(f_{k},F_{k})_{k=1:q_{3}}%
\end{array}%
,$ denotes the bivariate Fox's $H$-Function \cite{mittal}.
\end{proposition}

\begin{IEEEproof}
By involving\ (\ref{cdffinal1}) into (\ref{ora1}), and with the help of \cite%
[Eq. (07.34.03.0271.01)]{wolfram}, we obtain%
\begin{align}
\overline{C}_{ORA}& =\frac{\alpha _{2}}{2\ln (2)\Gamma (m_{1})\Gamma (\mu
_{2})}\sum_{n=0}^{m_{1}-1}\sum_{p=0}^{n}\left(
\begin{array}{c}
n \\
p%
\end{array}%
\right) \frac{\left( \frac{m_{1}}{\overline{\gamma }_{1}}\right) ^{n-p}}{n!}
\notag \\
& \times \int_{0}^{\infty }z^{n-p}e^{-\frac{m_{1}}{\overline{\gamma }_{1}}%
z}H_{1,1}^{1,1}\left( z\left\vert
\begin{array}{c}
(0,1);- \\
(0,1);-%
\end{array}%
\right. \right)  \notag \\
& \times H_{0,3}^{3,0}\left( \mu _{2}\left( \frac{m_{1}^{2}Cz}{\overline{%
\gamma }_{1}\overline{\gamma }_{2}^{(1)}}\right) ^{\frac{\alpha _{2}}{2}%
}\left\vert
\begin{array}{c}
-;- \\
\Delta _{1};-%
\end{array}%
\right. \right) dz.
\end{align}%
By making use of the identities \cite[Eq. (07.34.03.0228.01)]{wolfram} and
\cite[Eq. (2.3)]{mittal}, and using the same aforementioned steps on (\ref%
{cdffinal2}), we obtain the closed-form expression for the ACC under ORA
policy given in\ (\ref{orafinal}).
\end{IEEEproof}

It is worth mentioning that the univariate and bivariate Fox's $H$-Function
can be implemented efficiently in Matlab and Mathematica \cite{illi}.

\subsection{Optimal power and rate adaptation policy}

In the context of EH-enabled communications, several works have dealt with
optimal rate and/or power adaptation such as in \cite{kang, liu}. To this
end, the bandwidth-normalized ACC under OPRA policy is defined as the
capacity of a fading channel with the source transmit power being adapted to
maximize the achievable end-to-end capacity. Mathematically, it is expressed
as \cite{capacity}%
\begin{equation}
\overline{C}_{OPRA}=\int_{\gamma ^{\ast }}^{\infty }\log _{2}\left( \frac{z}{%
\gamma ^{\ast }}\right) f_{\gamma _{eq}}(z)dz,  \label{opra11}
\end{equation}%
which can be alternatively written as
\begin{equation}
\overline{C}_{OPRA}=\frac{1}{\ln (2)}\int_{\gamma ^{\ast }}^{\infty }\frac{1%
}{z}F_{\gamma _{eq}}^{c}(z)dz,  \label{opra1}
\end{equation}%
where $\gamma ^{\ast }$ is the optimal cutoff SNR, that is the zero of the
following function \cite{capacity}
\begin{eqnarray}
g(x) &=&\int_{x}^{\infty }\left( \frac{1}{x}-\frac{1}{z}\right) f_{\gamma
_{eq}}(z)dz-1  \notag \\
&=&\frac{1}{x}\left( 1-F_{\gamma _{eq}}(x)\right) -\mathbb{E}_{\gamma _{eq}}%
\left[ \frac{1}{\gamma _{eq}}\right] _{x}-1,  \label{gamstar}
\end{eqnarray}%
where
\begin{eqnarray}
\mathbb{E}_{\gamma _{eq}}\left[ \frac{1}{\gamma _{eq}}\right] _{x}
&=&\int_{x}^{\infty }\frac{1}{z}f_{\gamma _{eq}}(z)dz  \notag \\
&=&\sum\limits_{k=1}^{3}\sum_{i=1}^{2}T_{k}^{(i)}\left( x\right) ,
\label{mom1}
\end{eqnarray}%
and
\begin{align}
T_{1}^{(i)}\left( x\right) & =\frac{\left( i-1+(-1)^{i+1}F_{P_{E}}\left(
P_{B}\right) \right) \alpha _{2}}{2\Gamma ^{2-i}(m_{1})\Gamma (\mu _{2})}%
\sum_{n=0}^{m_{1}-1}\sum_{p=0}^{n}\left(
\begin{array}{c}
n \\
p%
\end{array}%
\right) \frac{1}{n!}  \notag \\
& \times M_{1,3-i+1}^{3-i+1,1}\left( \mu _{2}\left( \frac{m_{1}^{2-i}C}{%
\overline{\gamma }_{2}^{(i)}}\right) ^{\frac{\alpha _{2}}{2}}\left\vert
\begin{array}{c}
\eta \left( x\right) ;- \\
\Phi _{i};-%
\end{array}%
\right. \right) ,  \label{t1}
\end{align}%
\begin{align}
T_{2}^{(i)}\left( x\right) & =-\frac{\left( i-1+(-1)^{i+1}F_{P_{E}}\left(
P_{B}\right) \right) \alpha _{2}\left( n-p\right) }{2\Gamma
^{2-i}(m_{1})\Gamma (\mu _{2})}\sum_{n=0}^{m_{1}-1}\sum_{p=0}^{n}  \notag \\
& \times \frac{\left(
\begin{array}{c}
n \\
p%
\end{array}%
\right) }{n!}M_{1,3-i+1}^{3-i+1,1}\left( \mu _{2}\left( \frac{m_{1}^{2-i}C}{%
\overline{\gamma }_{2}^{(i)}}\right) ^{\frac{\alpha _{2}}{2}}\left\vert
\begin{array}{c}
\xi \left( x\right) ;- \\
\Phi _{i};-%
\end{array}%
\right. \right) ,  \label{t2}
\end{align}%
\begin{align}
T_{3}^{(i)}\left( x\right) & =-\frac{\left( i-1+(-1)^{i+1}F_{P_{E}}\left(
P_{B}\right) \right) \alpha _{2}^{2}}{4\Gamma ^{2-i}(m_{1})\Gamma (\mu _{2})}%
\sum_{n=0}^{m_{1}-1}\sum_{p=0}^{n}\frac{\left(
\begin{array}{c}
n \\
p%
\end{array}%
\right) }{n!}  \notag \\
& \times M_{2,4-i+1}^{3-i+1,2}\left( \mu _{2}\left( \frac{m_{1}^{2-i}C}{%
\overline{\gamma }_{2}^{(i)}}\right) ^{\frac{\alpha _{2}}{2}}\left\vert
\begin{array}{c}
(0,1,1),\xi \left( x\right) ;- \\
\Phi _{i};(1,1,1)%
\end{array}%
\right. \right) ,  \label{t3}
\end{align}%
with $\Phi _{1}=(\mu _{2},1,1),\left( m_{1}+\mu _{2}\left( 1-\frac{\alpha
_{2}}{2}\right) ,\frac{\alpha _{2}}{2},1\right) ,\left( p,\frac{\alpha _{2}}{%
2},1\right) ,$ $\Phi _{2}=(\mu _{2},1,1),\left( p,\frac{\alpha _{2}}{2}%
,1\right) ,$ $\eta \left( x\right) =\left( p-n,\frac{\alpha _{2}}{2},\frac{%
m_{1}}{\overline{\gamma }_{1}}x\right) ,$ and $\xi \left( x\right) =\left(
1+p-n,\frac{\alpha _{2}}{2},\frac{m_{1}}{\overline{\gamma }_{1}}x\right) .$

\begin{remark}
\bigskip One can notice clearly that by setting $x=0$ in (\ref{mom1}), the
moment of order $n=-1$ of $\gamma _{eq}$ is achieved. Consequently, the
incomplete Fox's $H$-Functions in (\ref{t1})-(\ref{t3}) become conventional
Fox's $H$-Functions.
\end{remark}

By differentiating (\ref{gamstar}) with respect to $x$, and computing its
limits at $0$ and $\infty $, one gets%
\begin{equation}
g^{\prime }(x)=-\frac{1}{x^{2}}F_{\gamma _{eq}}^{c}(x)<0,
\end{equation}%
\begin{equation}
\lim_{x\rightarrow 0}g\left( x\right) =\lim_{x\rightarrow 0}\int_{x}^{\infty
}\left( \frac{1}{x}-\frac{1}{z}\right) f_{\gamma _{eq}}(z)dz-1=\infty ,
\end{equation}

and%
\begin{equation}
\lim_{x\rightarrow \infty }g\left( x\right) =\lim_{x\rightarrow \infty
}\int_{x}^{\infty }\left( \frac{1}{x}-\frac{1}{z}\right) f_{\gamma
_{eq}}(z)dz-1=-1.
\end{equation}

Given the above, the zero value of (\ref{gamstar}) is unique.

By involving (\ref{cdffinal}) into (\ref{opra1}) and using Lemma 1 for $%
x=\gamma ^{\ast }$, a closed-form expression for the capacity under OPRA\
policy can be derived as%
\begin{align}
\overline{C}_{OPRA}& =\frac{\alpha _{2}}{2\ln (2)\Gamma (\mu _{2})}%
\sum_{n=0}^{m_{1}-1}\sum_{p=0}^{n}\frac{1}{n!}\left(
\begin{array}{c}
n \\
p%
\end{array}%
\right)  \notag \\
& \times \left[
\begin{array}{c}
\frac{F_{P_{E}}\left( P_{B}\right) }{\Gamma (m_{1})}M_{1,3}^{3,1}\left( \mu
_{2}\left( \frac{m_{1}C}{\overline{\gamma }_{2}^{(1)}}\right) ^{\frac{\alpha
_{2}}{2}}\left\vert
\begin{array}{c}
\xi \left( \gamma ^{\ast }\right) ;- \\
\Phi _{1};-%
\end{array}%
\right. \right) \\
+F_{P_{E}}^{c}\left( P_{B}\right) M_{1,2}^{2,1}\left( \mu _{2}\left( \frac{C%
}{\overline{\gamma }_{2}^{(2)}}\right) ^{\frac{\alpha _{2}}{2}}\left\vert
\begin{array}{c}
\xi \left( \gamma ^{\ast }\right) ;- \\
\Phi _{2};-%
\end{array}%
\right. \right)%
\end{array}%
\right] .  \label{oprafinal}
\end{align}

\subsection{Channel inversion with fixed rate\ Policy}

The CIFR policy is an adaptive transmission policy which requires that the
transmitter exploits the channel state information (CSI) {of both hops
(i.e., }$S$-$R$ and $R$-$D$ hops), so that a constant SNR is maintained at
the receiver (i.e., it inverts the channel fading). Mathematically speaking,
the bandwidth-normalized ACC under CIFR\ policy can be formulated as \cite%
{capacity}%
\begin{equation}
\overline{C}_{CIFR}=\log _{2}\left( 1+\frac{1}{\int_{0}^{\infty }\frac{%
f_{\gamma _{eq}}(z)}{z}dz}\right) .  \label{eqq}
\end{equation}

By setting $x=0$ in (\ref{mom1})-(\ref{t3}), the integral $\int_{0}^{\infty }%
\frac{f_{\gamma _{eq}}(z)}{z}dz$ is attained. Thus, the ACC\ under CIFR\
policy can be expressed as
\begin{equation}
\overline{C}_{CIFR}=\log _{2}\left( 1+\frac{1}{\sum\limits_{k=1}^{3}\sum%
\limits_{i=1}^{2}T_{k}^{(i)}\left( 0\right) }\right) .  \label{cifrfinal}
\end{equation}

\subsection{Truncated channel inversion with fixed rate policy}

As in the above-mentioned CIFR policy, TCIFR policy consists of channel
fading inversion only above a fixed cutoff $\gamma _{0}$. The ACC under this
policy is defined as
\begin{equation}
\overline{C}_{TCIFR}=\log _{2}\left( 1+\frac{1}{\int_{\gamma _{0}}^{\infty }%
\frac{f_{\gamma _{eq}}(z)}{z}dz}\right) \left( 1-F_{\gamma _{eq}}\left(
\gamma _{0}\right) \right) ,  \label{tcifr1}
\end{equation}

As done with\ the CIFR\ policy in (\ref{eqq}), by setting $x=\gamma _{0}$ in
(\ref{mom1})-(\ref{t3}), the integral $\int_{\gamma _{0}}^{\infty }\frac{%
f_{\gamma _{eq}}(z)}{z}dz$ is attained. Thus, the ACC\ under TCIFR\ policy
can be expressed as
\begin{equation}
\overline{C}_{TCIFR}=\left( 1-F_{\gamma _{eq}}\left( \gamma _{0}\right)
\right) \log _{2}\left( 1+\frac{1}{\sum\limits_{k=1}^{3}\sum%
\limits_{i=1}^{2}T_{k}^{(i)}\left( \gamma _{0}\right) }\right) .
\label{tcifrfinal}
\end{equation}

It is noteworthy that (\ref{aserfinal}), (\ref{orafinal}), (\ref{oprafinal}%
), (\ref{cifrfinal}), and (\ref{tcifrfinal}) are new, general and
easy-to-evaluate analytically for the ASER\ and ACC subject to various
transmit policies. We emphasize that such expressions can be served as a
benchmark for future studies, in addition to filling the gap that still
exists in the current technical literature.

\section{Numerical Results}

In this section, some illustrative numerical examples are depicted in order
to highlight the effects of key system parameters on the obtained
performance metrics. Without loss of generality, we set $m_{1}=3$, $\Omega
_{1}=\Omega _{2}=5,$ as the $S$-$R$ hop fading parameter and average power,
and $\alpha _{2}=2,$ for the $R$-$D$ hop fading parameters. In addition, the
relaying fixed-gain constant was set to $C=1$. Additionally, we set $%
T_{0}=T_{1}=1$s, and $\varepsilon =\varrho =\theta =0.7,$ and $B_{R}=500$ mAh%
$\times $V as the time slots in seconds, the harvester RF-to-DC\ conversion
efficiency, the time slot ratio, and the battery capacity, respectively.
Additionally, the complex contours of integration for computing (\ref%
{orafinal}) were chosen so as to separate left half-plane poles and right
half-plane ones, for both integrand terms on $s$ and $t$, that are $\mathcal{%
C}_{s}=\left] c_{s}-i\infty ,c_{s}+i\infty \right[ $ and $\mathcal{C}_{t}=%
\left] c_{t}-i\infty ,c_{t}+i\infty \right[ $, \ with $0<c_{s}<1$ and $%
c_{t}>\max \left( -\mu _{2},-\frac{2k}{\alpha _{2}},\frac{-2\mu _{2}}{\alpha
_{2}}\left( m_{1}+\mu _{2}\left( 1-\frac{\alpha _{2}}{2}\right) \right)
\right) .$

\begin{figure}[tbp]
\begin{center}
\hspace*{-1cm}\includegraphics[scale=.52]{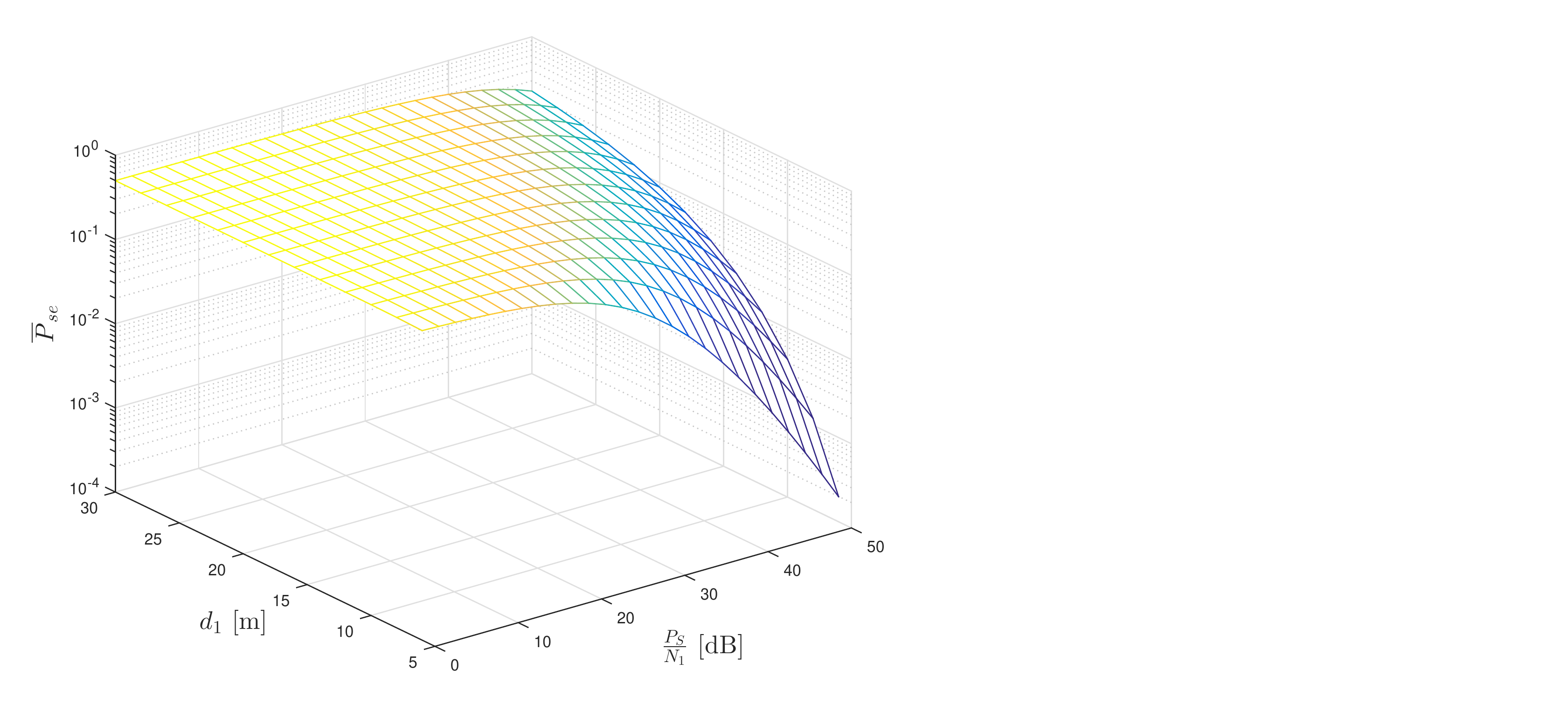}
\end{center}
\caption{ASER\ of the considered communication system for BPSK\ modulation
versus $\frac{P_{S}}{N_{1}}$ and $d_{1}.$}
\label{Fig:1}
\end{figure}

\begin{figure}[tbp]
\begin{center}
\includegraphics[scale=.7]{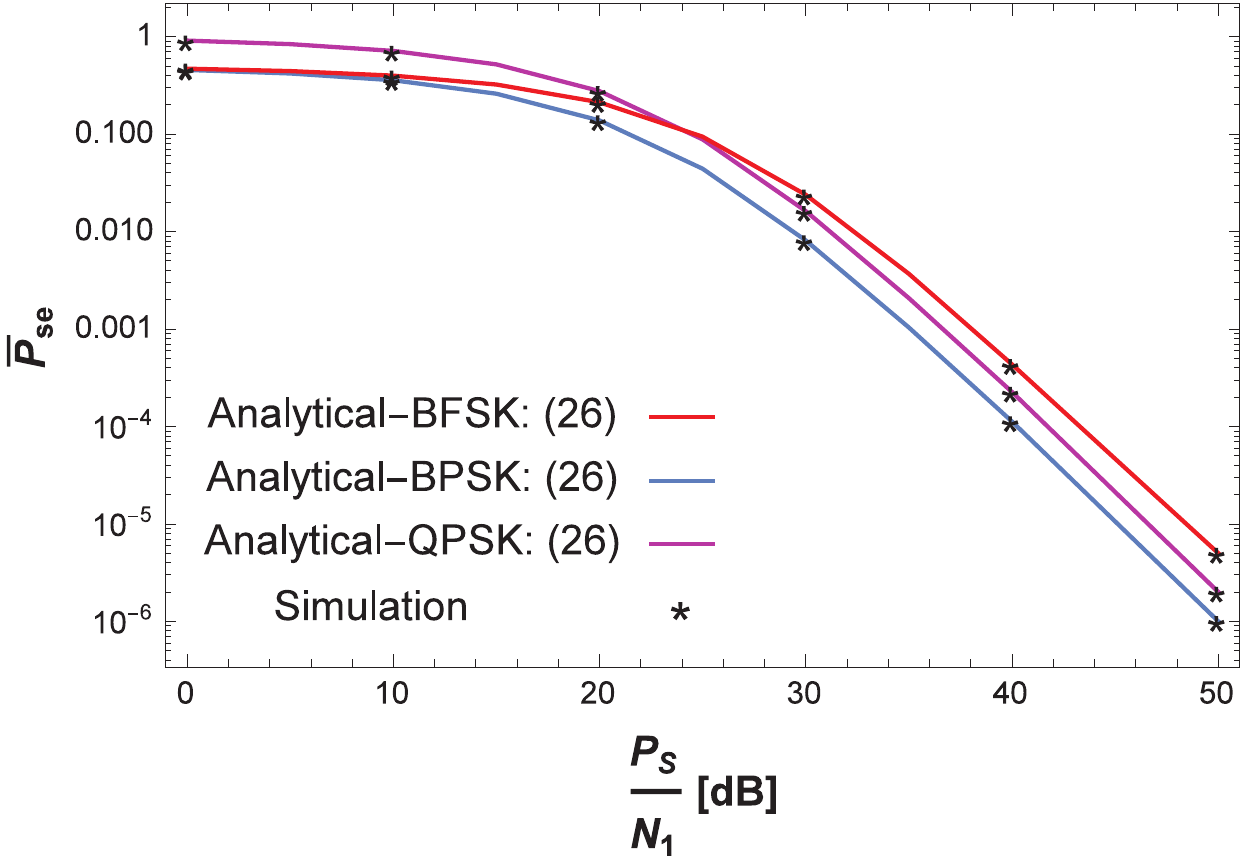}
\end{center}
\caption{ASER\ of the considered communication system versus $\frac{P_{S}}{%
N_{1}}$, for various modulation schemes.}
\label{Fig:2}
\end{figure}

Figs. \ref{Fig:1} and \ref{Fig:2} depict the ASER\ performance vs the
transmit power-to-noise ratio $\frac{P_{S}}{N_{1}}$ and the $S$-$R$ distance
$d_{1}$. We consider TS scheme, with $\frac{P_{S}}{N_{2}}=100$ dB, $d_{2}=25$
m, $m_{1}=3,$ and $\mu _{2}=4.2$, for various modulation schemes, based on (%
\ref{aserfinal}). The analytical and simulation results match, which
demonstrates the accuracy of the derived results. Additionally, it can be
clearly seen that the ASER decreases significantly as a function of $\frac{%
P_{S}}{N_{1}}$. In fact, the greater the transmit power, the higher is the
harvested energy, and consequently, the better is the bit error rate
performance. That is, for $\frac{P_{S}}{N_{1}}\rightarrow \infty ,$ we
converge to the performance of an AWGN\ channel (no fading). Furthermore,
one can remark also from Fig. \ref{Fig:1} the nodes distance effect on the
ASER\ performance. The farther $S$ and $R$, the more severe the path loss,
degrading the overall system's performance. Finally, the ASER\ performance
result shows that BPSK\ modulation scheme results in a better error rate
performance compared to its BFSK\ and QPSK counterparts.

\begin{figure}[tbp]
\begin{center}
\includegraphics[scale=.7]{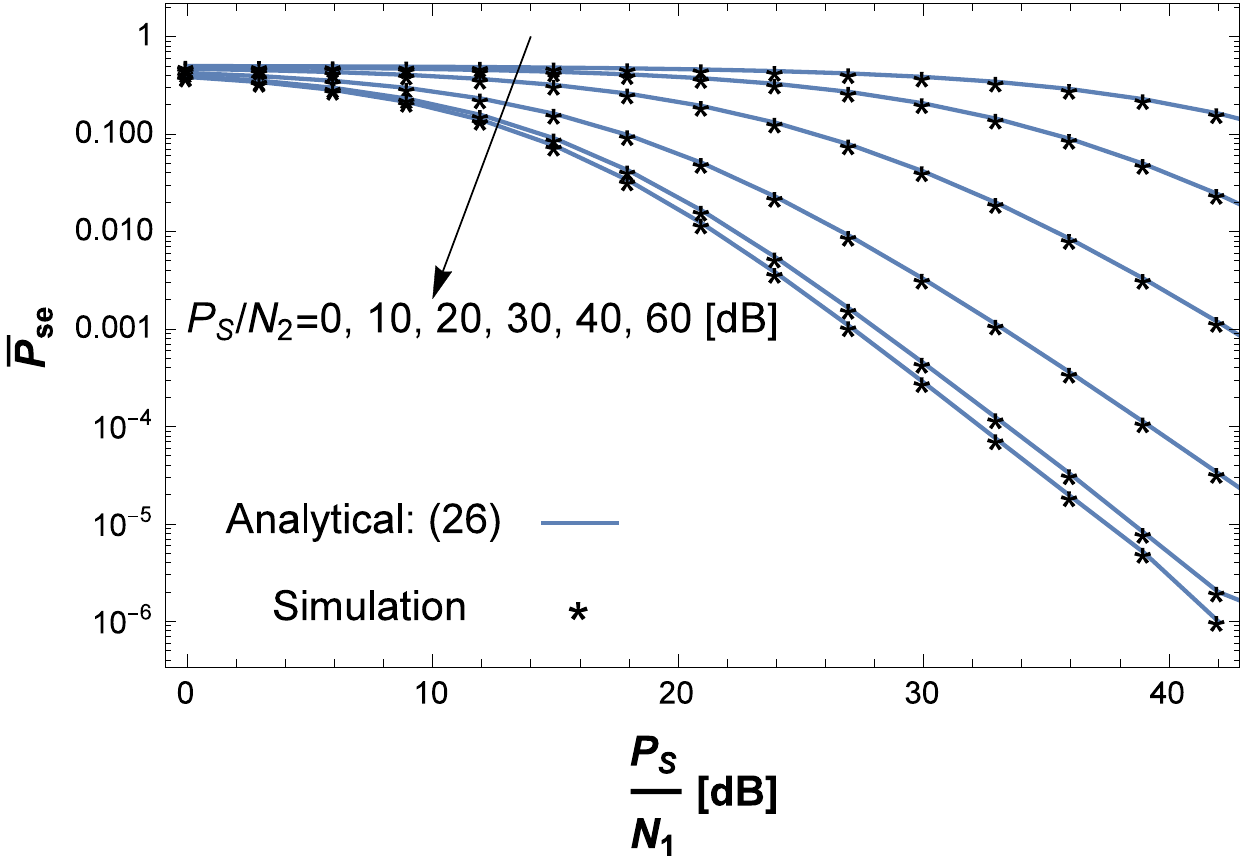}
\end{center}
\caption{ASER versus $\frac{P_{S}}{N_{1}}$ considering TS\ scheme, for
several $\frac{P_{S}}{N_{2}}$ values.}
\label{figg}
\end{figure}

Fig. \ref{figg} shows the ASER\ performance as a function of $\frac{P_{S}}{%
N_{1}}$, for several values of $\frac{P_{S}}{N_{2}}.$ Similar to Figs. \ref%
{Fig:1} and \ref{Fig:2}, it can be noticed that the ASER\ increases
significantly as a function of $\frac{P_{S}}{N_{1}}$. Interestingly, one can
remark that above certain $\frac{P_{S}}{N_{2}}$ values (e.g., $40$ dB), for
a given $\frac{P_{S}}{N_{1}}$ value, the ASER\ is not improved
significantly. That is, the end-to-end ASER\ performance converges to that
of the $S$-$R$ hop.

\begin{figure}[tbp]
\begin{center}
\includegraphics[scale=.7]{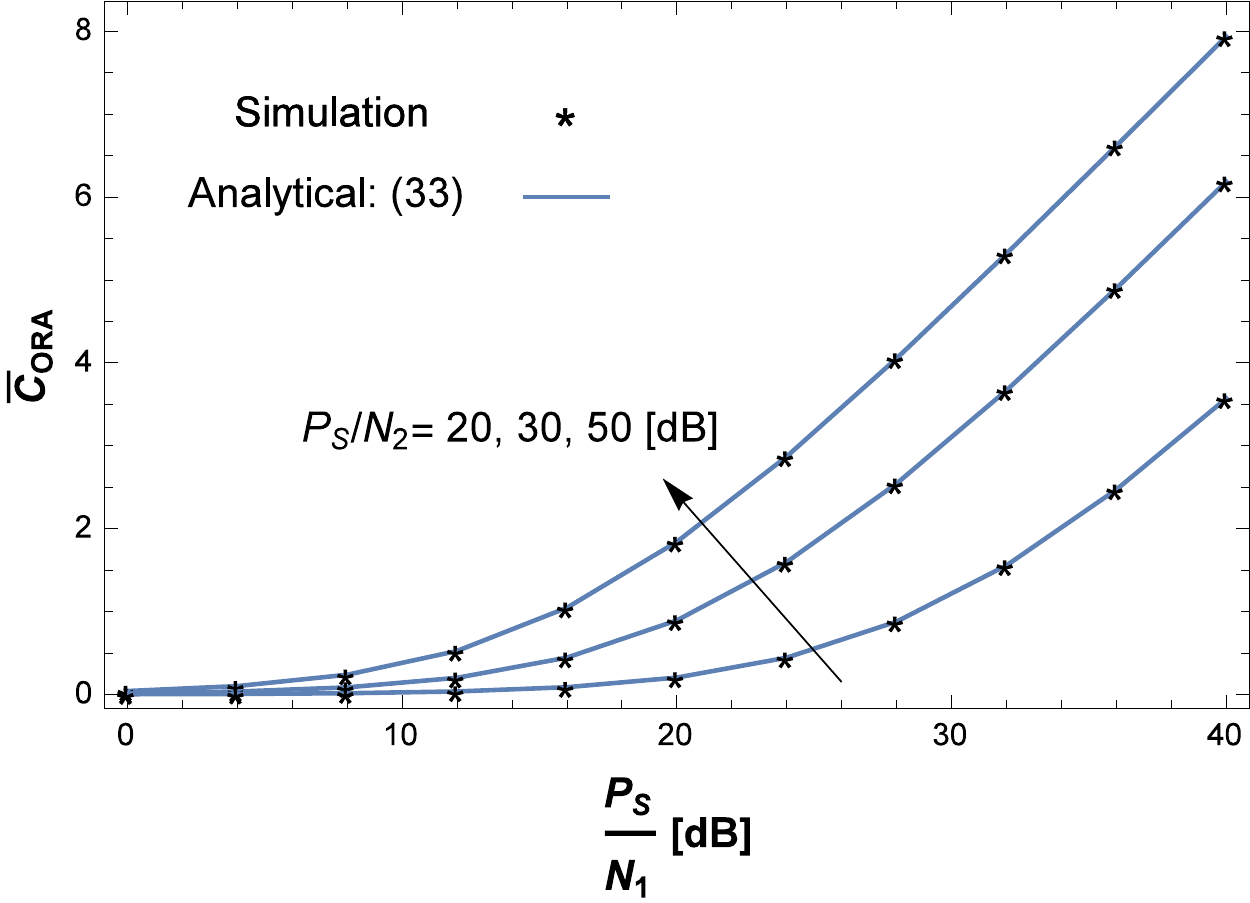}
\end{center}
\caption{ACC\ under ORA and OPRA policies\ of the considered communication
system for TS\ scheme versus $\frac{P_{S}}{N_{1}}$.}
\label{Fig:3}
\end{figure}

\bigskip Fig. \ref{Fig:3} shows the ACC\ performance under ORA\ and OPRA
policies, which are given by (\ref{orafinal}) and (\ref{oprafinal}),
respectively, versus $\frac{P_{S}}{N_{1}},$ with $\frac{P_{S}}{N_{2}}=100$
dB, $d_{1}=d_{2}=30$~m, and $\mu _{2}=4.2$. The respective optimal SNR\ $%
\gamma ^{\ast }$ for the OPRA\ policy was computed numerically using (\ref%
{gamstar}). Similarly to ASER, it can be evidently seen that the overall
ACC\ improves by increasing $\frac{P_{S}}{N_{1}}$, that is either greater
radiated power $P_{s}$ or lower noise power at the relay $N_{2}.$
Additionally, the result confirms again the free space path-loss effect, due
to the propagation distance, on the end-to-end system's performance. In
fact, the greater the distance $d_{i},$ $i=1,2$, the lower the received
end-to-end SNR, which results in degradation on the overall channel
capacity. Furthermore, it can be evidently seen that ACC under OPRA\ policy
outperforms the ORA\ one$,$ more particularly at high average SNR\ values.

\begin{figure}[tbp]
\begin{center}
\includegraphics[scale=.7]{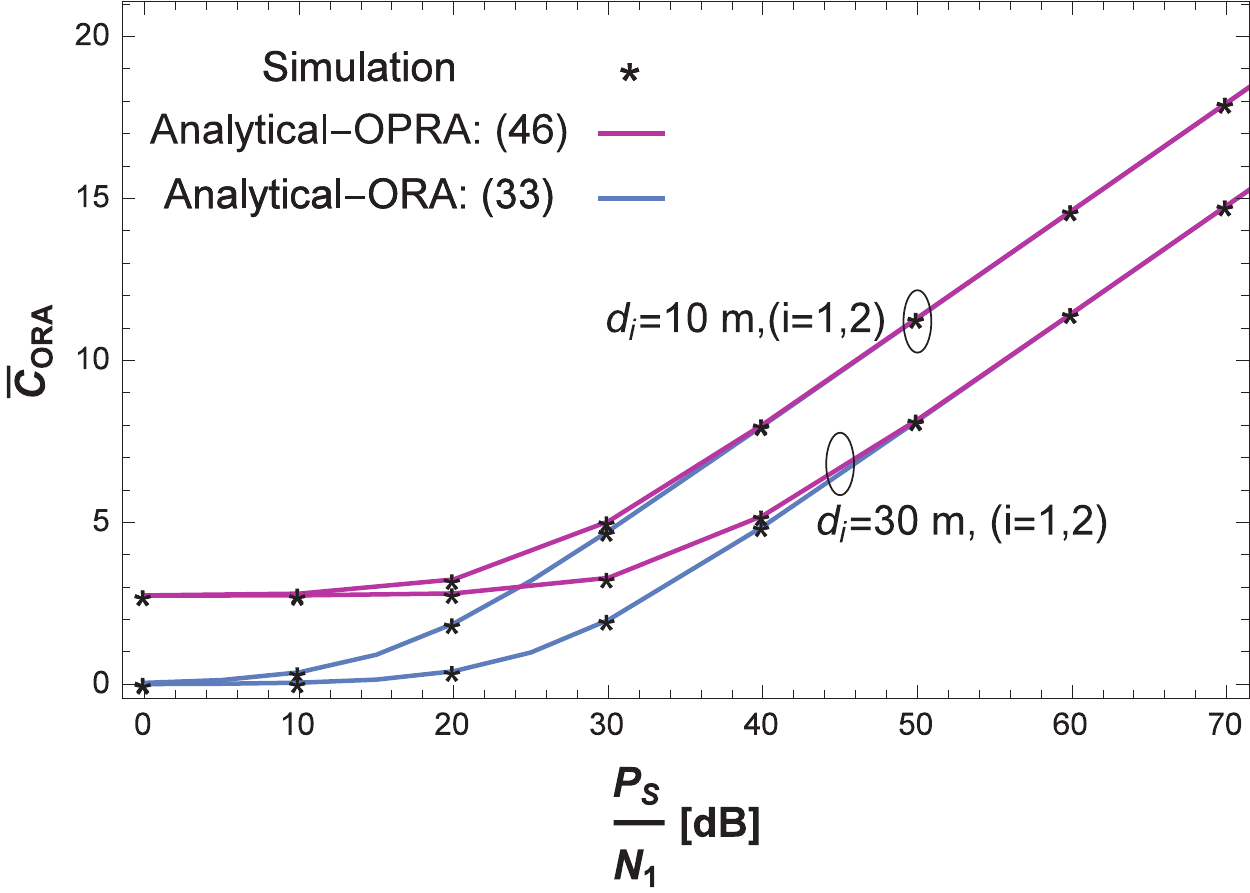}
\end{center}
\caption{ACC\ under ORA policiy\ of the considered communication system for
TS\ scheme versus $\frac{P_{S}}{N_{1}}$, for various $\frac{P_{S}}{N_{2}}$
values.}
\label{figbb}
\end{figure}

In Fig. \ref{figbb}, the ACC\ under ORA\ policy is shown versus $\frac{P_{S}%
}{N_{1}}$ for various values of $\frac{P_{S}}{N_{2}},$ for $d_{i}=10$ m.
Similarly to the $S$-$R$ power-to-noise ratio, the ACC\ increases also
significantly by increasing the value $\frac{P_{S}}{N_{2}}.$ Additionally,
at higher values of $\frac{P_{S}}{N_{2}}$ (e.g., beyond $50$ dB), the ACC\
becomes steady for a given $\frac{P_{S}}{N_{1}}$ value. That is, the
end-to-end SNR\ converges to the $S$-$R$ SNR, and consequently, the
end-to-end ACC\ is that of the $S$-$R$ link.

\begin{figure}[tbp]
\begin{center}
\hspace*{-1cm}\includegraphics[scale=.52]{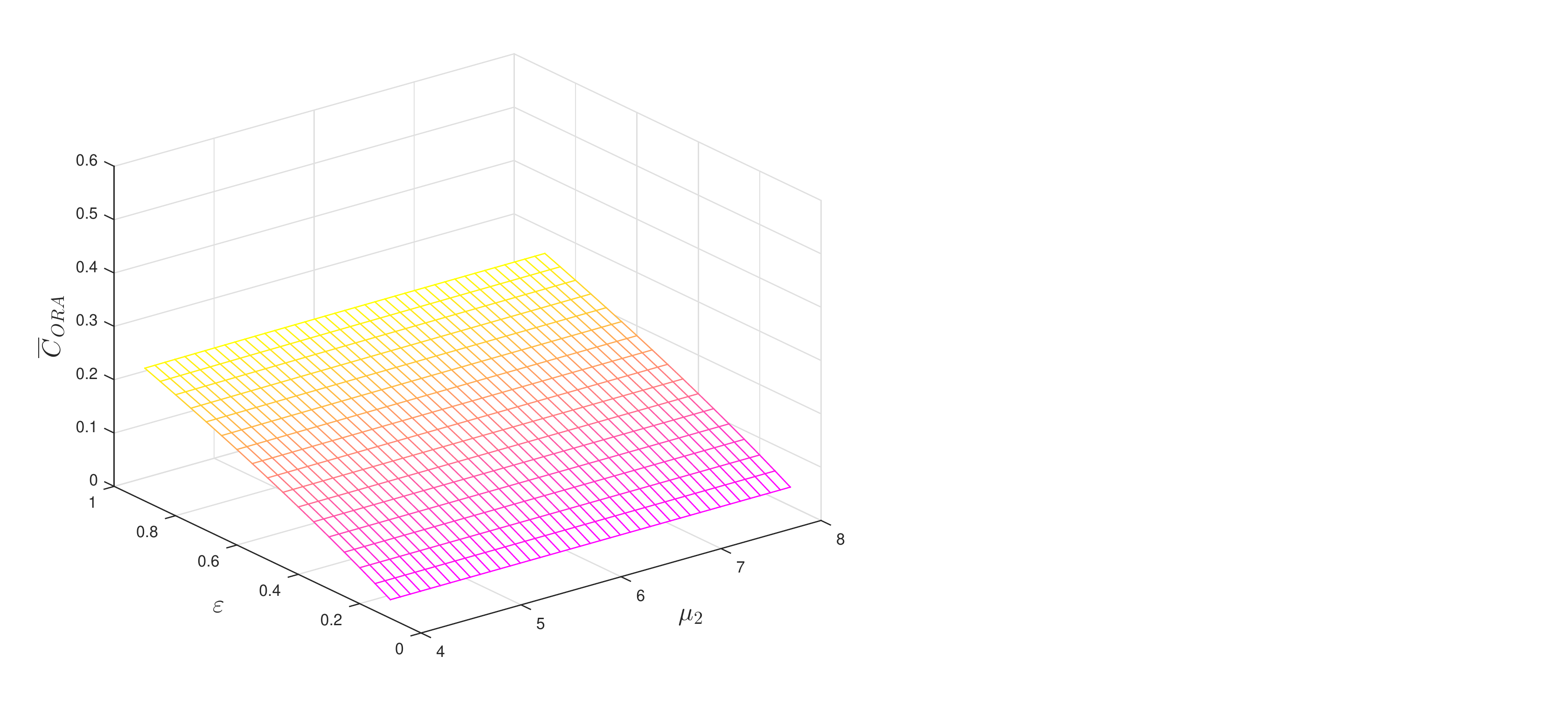}
\end{center}
\caption{ACC\ under ORA policy\ of the considered communication system for
TS\ scheme versus $\protect\varepsilon $ and $\protect\mu _{2}$.}
\label{Fig:4}
\end{figure}

In fig. \ref{Fig:4}, the ACC\ under ORA\ policy is shown versus the
harvester efficiency $\varepsilon $ as well as the fading parameter $\mu
_{2},$ for $\frac{P_{S}}{N_{0i}}=20$ dB, $d_{i}=10$ m, $i=1,2.$ One can
ascertain that the ACC\ increases slightly as a function of the harvester
efficiency $\varepsilon .$ This shows that the greater $\varepsilon $, the
higher the harvested power $P_{R},$ and consequently the relay can forward
the information signal with a higher power. Interestingly, as we use TS\
protocol, even by increasing the duration dedicated to energy harvesting
over information decoding one, the received SNR\ at $R$ is not impacted as
can be seen from (\ref{gam1}). In addition to this, the results show that
the ACC\ is not impacted by the fading parameter $\mu _{2}$

\begin{figure}[tbp]
\begin{center}
\hspace*{-1cm}\includegraphics[scale=.52]{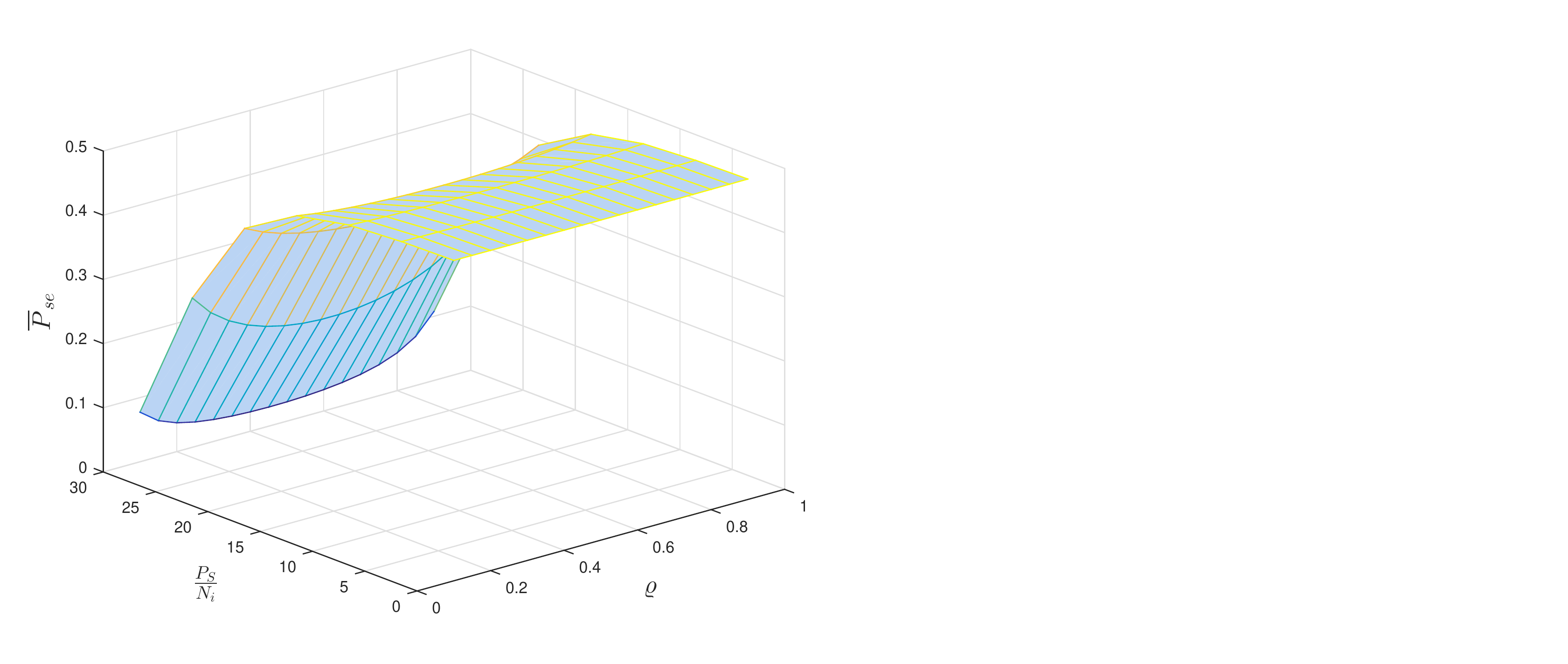}
\end{center}
\caption{ASER\ of the considered communication system for PS\ scheme versus $%
\protect\varrho $ and $\frac{P_{S}}{N_{i}},i=1,2$.}
\label{figps1}
\end{figure}

Fig. \ref{figps1} highlights the ASER\ evolution versus the average SNR\ $%
\frac{P_{S}}{N_{i}},i=1,2$, as well as the ratio $\varrho $, for PS\ scheme.
Differently from the TS\ scheme, one can notice evidently that the ASER\
evolution versus $\varrho $ admits a minimum, particularly for high average
SNR\ values, while for TS\ scheme as in fig. \ref{Fig:4}, the capacity
increases as a function of $\varepsilon $. This can be explained from (\ref%
{gam1}) and (\ref{snr1ps}), where the SNR\ of the first hop for the PS\
scheme is affected by the ratio $\varrho ,$ whereas for its TS\ counterpart,
the SNR\ $\gamma _{1}$ is not impacted by the ratio $\varepsilon .$

\begin{figure}[tbp]
\begin{center}
\includegraphics[scale=.7]{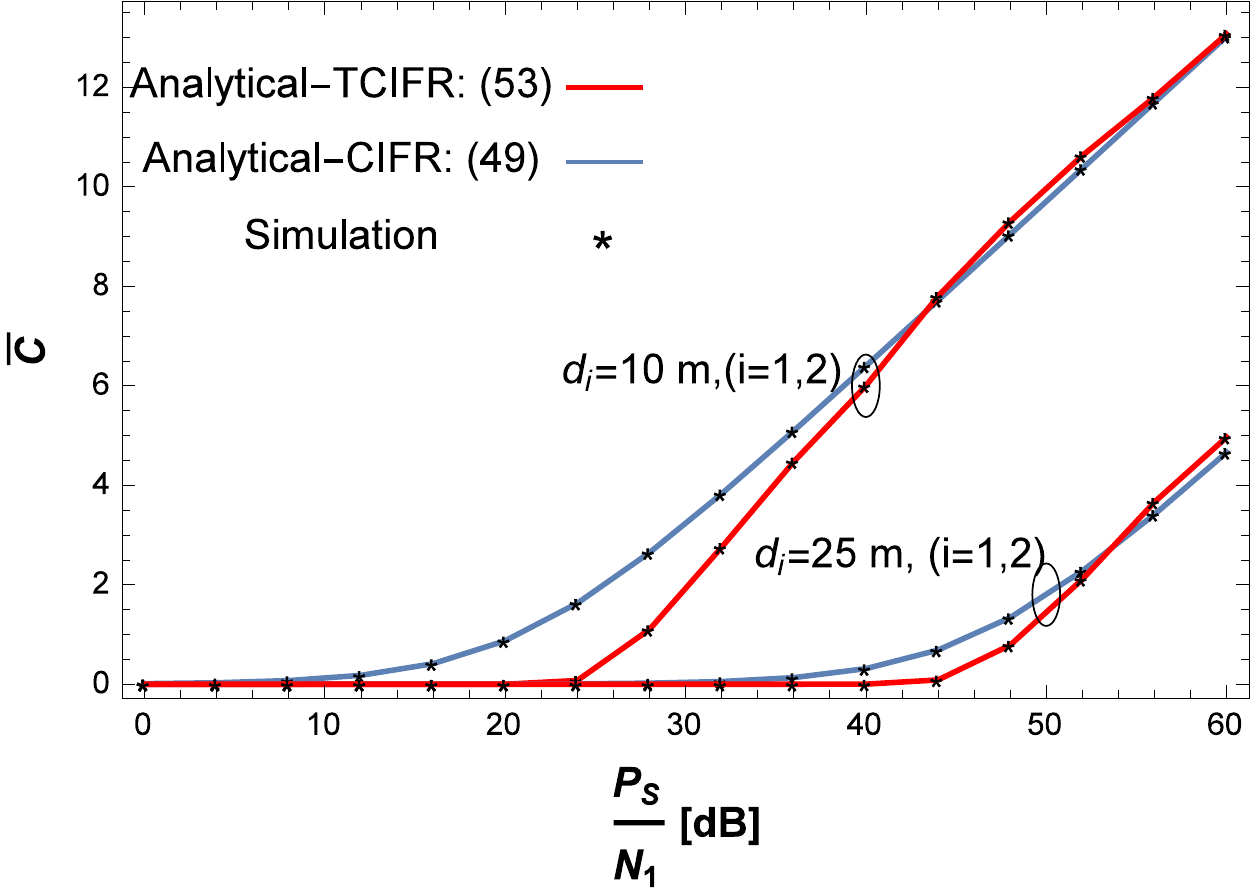}
\end{center}
\caption{ACC\ of the considered communication system under CIFR\ and TCIFR
policies versus $\frac{P_{S}}{N_{1}},$ with $\frac{P_{S}}{N_{2}}=30$ dB.}
\label{cifr1}
\end{figure}

In Fig. \ref{cifr1}, the ACC under CIFR\ and TCIFR policies\ is plotted
versus the average SNR\ $\frac{P_{S}}{N_{1}},i=1$, with $\frac{P_{s}}{N_{2}}%
=30$ dB, for TS\ and PS\ schemes. One can remark clearly that the ACC\ under
CIFR\ policy surpasses slightly that under TCIFR policy, particularly for
average SNR\ values that are less than $25$ dB. Nevertheless, this behavior
is changed at high SNR\ values, where the TCIFR\ capacity slightly
outperforms the CIFR\ one. In fact, in high average SNR regime, the
end-to-end SNR\ is relatively high. That is, the truncation of the channel
inversion does not improve significantly the end-to-end channel capacity.
Furthermore, the greater the $S$-$R$ and $R$-$D$ distance, the higher the
free-space path loss.

\begin{figure}[tbp]
\begin{center}
\includegraphics[scale=.7]{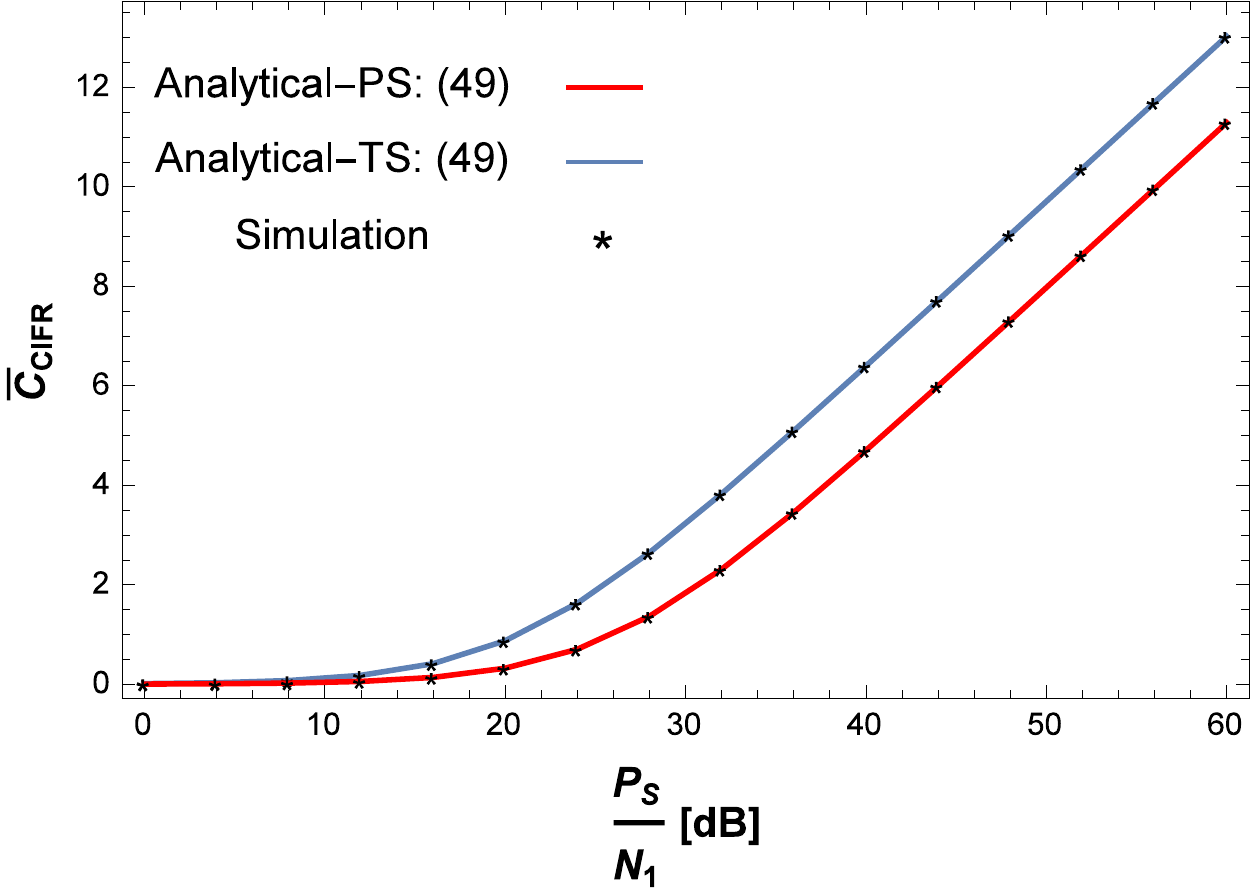}
\end{center}
\caption{ACC\ of the considered communication system under CIFR\ policy
versus $\frac{P_{S}}{N_{1}},$ with $\frac{P_{S}}{N_{2}}=30$ dB, $d_i$=10 m,
for TS\ and PS\ schemes.}
\label{cifr2}
\end{figure}

The ACC\ under CIFR\ policy is plotted for both TS\ and PS\ schemes in Fig. %
\ref{cifr2}. Interestingly, one can ascertain from the curves that the TS
protocol clearly outperforms its PS\ counterpart in terms of capacity
performance.

\section{Conclusion}

In this paper, a performance analysis of an AF\ dual-hop energy
harvesting-based wireless communication system subject to asymmetric fading
channels namely, Nakagami-$m,$ and $\alpha $-$\mu $ was carried out. Both
SWIPT\ schemes, namely power splitting and time switching, have been
considered in the analysis. Statistical properties such as the CDF and PDF
of the end-to-end SNR were retrieved, in generalized expressions for both
TS\ and PS\ schemes, based on which, closed-form expressions of the ASER as
well as ACC over ORA, OPRA, CIFR, and TCIFR adaptive transmission policies
have been derived. These analytical results were corroborated by Monte-Carlo
simulation.

\section*{Appendix A: Proof of Proposition 1}

Given the end-to-end SNR\ expression in (\ref{mn}), the respective CDF\ of $%
\gamma _{eq}$ is defined as \cite{ansarithesis}
\begin{eqnarray}
F_{\gamma _{eq}}\left( z\right) &=&\Pr \left[ \gamma _{1}<z\left( 1+\frac{C}{%
\gamma _{2}}\right) \right]  \notag \\
&=&\int_{0}^{\infty }F_{\gamma _{1}}\left( z\left( 1+\frac{C}{x}\right)
\right) f_{\gamma _{2}}\left( x\right) dx.  \label{cdftot1}
\end{eqnarray}

Since $\gamma _{2}$ depends on the harvested power $P_{R},$ depending on
whether $E_{R}<B_{R}$ or $E_{R}\geq B_{R},$ the CDF\ of the end-to-end SNR\
can be expressed as
\begin{eqnarray}
F_{\gamma _{eq}}(z) &=&F_{\gamma _{eq}}(z\left\vert P_{E}<P_{B}\right. )\Pr
\left[ P_{E}<P_{B}\right]  \notag \\
&&+F_{\gamma _{eq}}(z\left\vert P_{E}\geq P_{B}\right. )\Pr \left[ P_{E}\geq
P_{B}\right] .
\end{eqnarray}

The conditional PDF\ of $\gamma _{eq}$ in each term of the above equation is
expressed as%
\begin{align}
F_{\gamma _{eq}}(z\left\vert P_{E}<P_{B}\right. )& =\underset{F_{\gamma
_{eq}^{(1)}}(z)}{\underbrace{\int_{0}^{\infty }F_{\gamma _{1}}\left( z\left(
1+\frac{C}{x}\right) \right) f_{\gamma _{2}^{(1)}}\left( x\right) dx}},
\label{51} \\
F_{\gamma _{eq}}(z\left\vert P_{E}\geq P_{B}\right. )& =\underset{F_{\gamma
_{eq}^{(2)}}(z)}{\underbrace{\int_{0}^{\infty }F_{\gamma _{1}}\left( z\left(
1+\frac{C}{x}\right) \right) f_{\gamma _{2}^{(2)}}\left( x\right) dx}}.
\label{52}
\end{align}

One can note that evaluating (\ref{51}) and (\ref{52}) corresponds to two
distinct cases, namely $E_{R}<B_{R}$ and $E_{R}\geq B_{R},$ respectively.

\begin{itemize}
\item First case: $E_{R}<B_{R}$
\end{itemize}

One can notice from (\ref{gam2}) that the $\gamma _{2}^{(1)}$ is the product
of two random variables; namely $P_{E}$ and $\Upsilon _{2}.$\ Consequently,
its PDF\ is expressed as%
\begin{equation}
f_{\gamma _{2}^{(1)}}\left( z\right) =\int_{0}^{\infty }\frac{1}{t}%
f_{P_{E}}\left( \frac{z}{t}\right) f_{\Upsilon _{2}}\left( t\right) dt.
\end{equation}

By involving (\ref{pdfpwr}) and (\ref{gggt}) into the above equation, as
well as using the identity Eq. (07.34.03.0228.01) of \cite{wolfram}
alongside with some algebraic manipulations, one obtains{\small
\begin{align}
f_{\gamma _{2}^{(1)}}\left( z\right) & =\frac{\alpha _{2}\mu _{2}^{\mu _{2}}%
\overline{\gamma }_{2}^{(1)}\left( \frac{\Psi z}{\overline{\gamma }_{2}^{(1)}%
}\right) ^{\frac{\alpha _{2}\mu _{2}}{2}-1}}{2\Gamma (m_{1})\Gamma (\mu
_{2})z}\int_{0}^{\infty }G_{0,1}^{1,0}\left( \frac{\Psi z}{t}\left\vert
\begin{array}{c}
-;- \\
\omega ;-%
\end{array}%
\right. \right)  \notag \\
& \times G_{0,1}^{1,0}\left( \mu _{2}\left( \frac{t}{\overline{\Upsilon }_{2}%
}\right) ^{\frac{\alpha _{2}}{2}}\left\vert
\begin{array}{c}
-;- \\
0;-%
\end{array}%
\right. \right) dt.  \label{26}
\end{align}%
} where $G_{p,q}^{m,n}\left( .\left\vert .\right. \right) $ is the Meijer's $%
G$-function \cite[Eq. (07.34.02.0001.01)]{wolfram}, $\Psi $ is defined in (%
\ref{psii}), and $\omega =m_{1}+1-\frac{\alpha _{2}\mu _{2}}{2}.$
Furthermore, using \cite[Eqs. (07.34.16.0002.01, 07.34.21.0012.01)]{wolfram}
and with some further manipulations, (\ref{26}) can be expressed by
\begin{equation}
f_{\gamma _{2}^{(1)}}\left( z\right) =\frac{\alpha _{2}}{2z\Gamma
(m_{1})\Gamma (\mu _{2})}H_{0,2}^{2,0}\left( \mu _{2}\left( \frac{m_{1}z}{%
\overline{\gamma }_{2}^{(1)}}\right) ^{\frac{\alpha _{2}}{2}}\left\vert
\begin{array}{c}
-;- \\
\Upsilon ;-%
\end{array}%
\right. \right) .  \label{pdfgam2}
\end{equation}%
with $\Upsilon =(\mu _{2},1),\left( m_{1}+\mu _{2}\left( 1-\frac{\alpha _{2}%
}{2}\right) ,\frac{\alpha _{2}}{2}\right) .$

Substituting (\ref{pdfgam2}) and (\ref{cdfsx}) into (\ref{cdftot1}), and
using the finite sum representation of the lower incomplete gamma function
in \cite[Eq. (8.352.1)]{integrals} as well as \cite[Eq. (2.9.4)]{kilbas}
yields
\begin{eqnarray}
F_{\gamma _{eq}^{(1)}}\left( z\right) &=&1-\frac{\alpha _{2}e^{-\frac{m_{1}}{%
\overline{\gamma }_{1}}z}}{2\Gamma (m_{1})\Gamma (\mu _{2})}%
\sum_{n=0}^{m_{1}-1}\sum_{p=0}^{n}\left(
\begin{array}{c}
n \\
p%
\end{array}%
\right) \frac{\left( \frac{m_{1}}{\overline{\gamma }_{1}}z\right) ^{n-p}}{n!}
\notag \\
&&\times \int_{0}^{\infty }x^{-1}H_{0,1}^{1,0}\left( \frac{m_{1}Cz}{%
\overline{\gamma }_{1}x}\left\vert
\begin{array}{c}
-;- \\
\left( p,1\right) ;-%
\end{array}%
\right. \right)  \notag \\
&&\times H_{0,2}^{2,0}\left( \mu _{2}\left( \frac{m_{1}x}{\overline{\gamma }%
_{2}^{(1)}}\right) ^{\frac{\alpha _{2}}{2}}\left\vert
\begin{array}{c}
-;- \\
\Upsilon ;-%
\end{array}%
\right. \right) dx.
\end{eqnarray}

Finally, by applying \cite[Eqs. (2.1.3, 2.8.4)]{kilbas}, (\ref{cdffinal1})
is attained.

\begin{itemize}
\item Second case: $E_{R}\geq B_{R}$
\end{itemize}

In this scenario, the SNR $\gamma _{2}$ is expressed as in\ (\ref{snr22}).
Hence, its respective PDF can be derived from (\ref{gggt}) using \cite[Eq.
(2.9.4)]{kilbas} as%
\begin{align}
f_{\gamma _{2}^{(2)}}(z)& =\frac{\alpha _{2}\mu _{2}^{\mu _{2}}}{2\Gamma
(\mu _{2})\overline{\gamma }_{2}^{(2)}}\left( \frac{z}{\overline{\gamma }%
_{2}^{(2)}}\right) ^{\frac{\alpha _{2}\mu _{2}}{2}-1}e^{-\mu _{2}\left(
\frac{z}{\overline{\gamma }_{2}^{(2)}}\right) ^{\frac{\alpha _{2}}{2}}},
\notag \\
& =\frac{\alpha _{2}}{2\Gamma (\mu _{2})z}H_{0,1}^{1,0}\left( \mu _{2}\left(
\frac{z}{\overline{\gamma }_{2}^{(2)}}\right) ^{\frac{\alpha _{2}}{2}%
}\left\vert
\begin{array}{c}
-;- \\
\left( \mu _{2},1\right) ;-%
\end{array}%
\right. \right) .
\end{align}

Similarly to $F_{\gamma _{eq}}^{(1)}\left( z\right) ,$ by involving the PDF\
of $\gamma _{2}$ above and (\ref{cdfsx}) into (\ref{cdftot1}), and using
\cite[Eq. (8.352.1)]{integrals} as well as \cite[Eq. (2.9.4)]{kilbas} we
obtain%
\begin{align}
F_{\gamma _{eq}^{(2)}}\left( z\right) & =1-\frac{\alpha _{2}}{2\Gamma (\mu
_{2})}e^{-\frac{m_{1}}{\overline{\gamma }_{1}}z}\sum_{n=0}^{m_{1}-1}%
\sum_{p=0}^{n}\left(
\begin{array}{c}
n \\
p%
\end{array}%
\right) \frac{\left( \frac{m_{1}}{\overline{\gamma }_{1}}z\right) ^{n-p}}{n!}
\notag \\
& \times \int_{0}^{\infty }x^{-1}H_{1,0}^{0,1}\left( \frac{\overline{\gamma }%
_{1}x}{m_{1}Cz}\left\vert
\begin{array}{c}
(1-p,1);- \\
-;-%
\end{array}%
\right. \right) \\
& \times H_{0,1}^{1,0}\left( \mu _{2}\left( \frac{x}{\overline{\gamma }%
_{2}^{(2)}}\right) ^{\frac{\alpha _{2}}{2}}\left\vert
\begin{array}{c}
-;- \\
(\mu _{2},1);-%
\end{array}%
\right. \right) dx.
\end{align}

By applying \cite[Eqs. (2.1.3, 2.8.4)]{kilbas}, (\ref{cdffinal2}) is
obtained.

\section*{Appendix B: Proof of Corollary 1}

One can see that differentiating (\ref{cdffinal}) with respect to $z$ yields
\begin{equation}
f_{\gamma _{eq}}(z)=f_{\gamma _{eq}^{(1)}}(z)F_{P_{E}}\left( P_{B}\right)
+f_{\gamma _{eq}^{(2)}}(z)F_{P_{E}}^{c}\left( P_{B}\right) ,  \label{pdfstp1}
\end{equation}%
\begin{figure*}[t]
{\normalsize 
\setcounter{mytempeqncnt}{\value{equation}}
\setcounter{equation}{60} }
\par
\begin{equation}
f_{\gamma _{eq}^{(1)}}(z)=-\frac{\alpha _{2}e^{-\frac{m_{1}}{\overline{%
\gamma }_{1}}z}F_{P_{E}}\left( P_{B}\right) }{2\Gamma (m_{1})\Gamma (\mu
_{2})}\sum_{n=0}^{m_{1}-1}\sum_{p=0}^{n}\left(
\begin{array}{c}
n \\
p%
\end{array}%
\right) \frac{\left( \frac{m_{1}}{\overline{\gamma }_{1}}z\right) ^{n-p}}{n!}%
\left[
\begin{array}{l}
\left[ \frac{n-p}{z}-\frac{m_{1}}{\overline{\gamma }_{1}}\right]
H_{0,3}^{3,0}\left( \mu _{2}\left( \frac{m_{1}^{2}Cz}{\overline{\gamma }_{1}%
\overline{\gamma }_{2}^{(1)}}\right) ^{\frac{\alpha _{2}}{2}}\left\vert
\begin{array}{c}
-;- \\
\Delta _{1};-%
\end{array}%
\right. \right) \\
+\underset{\mathcal{J}}{\underbrace{\frac{\partial \left[ H_{0,3}^{3,0}%
\left( \mu _{2}\left( \frac{m_{1}^{2}Cz}{\overline{\gamma }_{1}\overline{%
\gamma }_{2}^{(1)}}\right) ^{\frac{\alpha _{2}}{2}}\left\vert
\begin{array}{c}
-;- \\
\Delta _{1};-%
\end{array}%
\right. \right) \right] }{\partial z}}}%
\end{array}%
\right]  \label{stpp1}
\end{equation}%
\par
{\normalsize 
\hrulefill 
\vspace*{4pt} }
\end{figure*}

By applying the derivative with respect to $z$ to (\ref{cdffinal1}), one
obtains (\ref{stpp1}) at the top of the next page, where $\Delta _{1}$ is
defined in Proposition 1. Using the definition of Fox's $H$-Function in \cite%
[Eqs. (1.5-1.6)]{yakub} and given that $-s=\frac{\Gamma \left( 1-s\right) }{%
\Gamma \left( -s\right) },$ by applying the derivative with respect to $z$,
one gets%
\begin{align}
\mathcal{J}& =-\frac{1}{2\pi j}\frac{\alpha _{2}}{2z}\int_{\mathcal{C}%
_{s}}\Gamma \left( \mu _{2}+s\right) \Gamma \left( m_{1}+\mu _{2}\left( 1-%
\frac{\alpha _{2}}{2}\right) +\frac{\alpha _{2}}{2}s\right)  \notag \\
& \times s\Gamma \left( p+\frac{\alpha _{2}}{2}s\right) \left( \mu
_{2}\left( \frac{m_{1}^{2}Cz}{\overline{\gamma }_{1}\overline{\gamma }%
_{2}^{(1)}}\right) ^{\frac{\alpha _{2}}{2}}\right) ^{-s}ds  \notag \\
& =H_{1,4}^{3,1}\left( \mu _{2}\left( \frac{m_{1}^{2}Cz}{\overline{\gamma }%
_{1}\overline{\gamma }_{2}^{(1)}}\right) ^{\frac{\alpha _{2}}{2}}\left\vert
\begin{array}{c}
(0,1);- \\
\Delta _{1};(1,1)%
\end{array}%
\right. \right) .  \label{35}
\end{align}

Substituting (\ref{35}) into (\ref{stpp1}), and performing in a similar way
the differentiation with respect to $z$ done in (\ref{35}) for (\ref%
{cdffinal2}), (\ref{pdffinal}) is attained.

\section*{Appendix C: Proof of Lemma 1}

By plugging the definition of Fox's $H$-Function \cite[Eqs. (1.2)-(1.3)]%
{mathai} into (\ref{mypsi}), we obtain
\begin{align}
\Xi (x)& =\frac{1}{2\pi j}\int_{\mathcal{L}}\frac{\prod\limits_{i=1}^{n}%
\Gamma \left( 1-a_{i}-A_{i}s\right) \prod\limits_{k=1}^{m}\Gamma \left(
b_{k}+B_{k}s\right) }{\prod\limits_{i=n+1}^{p}\Gamma \left(
a_{i}+A_{i}s\right) \prod\limits_{k=m+1}^{q}\Gamma \left(
1-b_{k}-B_{k}s\right) }  \notag \\
& \times a^{-s}ds\int_{x}^{\infty }e^{-\lambda z}z^{-vs+\omega -1}dz,
\label{ps2}
\end{align}%
where $\mathcal{L}$ is a complex contour of integration. Using \cite[Eq.
(8.350.2)]{integrals} and after some algebraic manipulations, we obtain
\begin{equation}
\int_{x}^{\infty }e^{-\lambda z}z^{-vs+\omega -1}dz=\lambda ^{-\omega
}\Gamma \left( \omega -vs,\lambda x\right) .  \label{intgg}
\end{equation}

Finally, by inserting (\ref{intgg}) into (\ref{ps2}), Lemma 1 is attained.

%
\bibliographystyle{IEEEtran}
\bibliography{references}

\end{document}